\ifx\XeTeXversion\undefined\pdfoutput=1\fi
\ifdefined\UseOneColumnSI
  \documentclass[11pt,twoside,notitlepage]{article}
\else
  \documentclass{article}
\fi

\usepackage{silence}
\usepackage[verbose=silent]{microtype}
\usepackage{graphicx}
\usepackage{array}
\usepackage{ragged2e}
\usepackage{booktabs}
\usepackage{xurl}
\usepackage{xr-hyper}
\usepackage{hyperref}
\ifdefined\UseOneColumnSI
  \usepackage[T1]{fontenc}
  \usepackage{lmodern}
  \usepackage[numbers,super,sort&compress]{natbib}
  \setcitestyle{numbers,super,comma,open={},close={}}
  \usepackage{si-one-column}
\else
  \PassOptionsToPackage{numbers,super,sort&compress}{natbib}
  \usepackage[preprint]{icml2026}
  \setcitestyle{numbers,super,comma,open={},close={}}
\fi
\AtBeginEnvironment{thebibliography}{\raggedright}

\usepackage{amsmath,amssymb}
\usepackage{amsthm}
\usepackage{mathtools}
\allowdisplaybreaks

\theoremstyle{plain}

\theoremstyle{definition}

\theoremstyle{remark}

\usepackage{multirow}
\usepackage{xcolor}
\usepackage{algorithm}
\usepackage{algorithmic}
\usepackage{etoolbox}
\usepackage{placeins}

\definecolor{opblue}{rgb}{0.13,0.33,0.62}

\definecolor{alggray}{gray}{0.95}
\makeatletter
\newcommand{\FORR}[1]{\ALC@it\algorithmicfor\ #1%
  \begin{ALC@loop}}
\newcommand{\ENDFORR}{\end{ALC@loop}\ALC@it\algorithmicendfor}
\newcommand{\RETURN}{\ALC@it\textbf{return}\ }

\AtBeginDocument{%
  \renewcommand{\algorithmiccomment}[1]{%
    \\
    {\raggedright\footnotesize\color{black!55}$\triangleright$~#1\par}%
  }%
}
\AtBeginEnvironment{algorithmic}{\raggedright}
\makeatother

\newcommand{\gopes}{\gamma_{\mathrm{wt}}}

\usepackage{bm}
\usepackage{tikz}\usetikzlibrary{arrows.meta,calc,decorations.pathreplacing,shapes.geometric,patterns}
\DeclareRobustCommand{\stmark}[1]{\tikz[baseline=(mk.base)]\node[circle,draw,line width=0.35pt,inner sep=0.15pt,minimum size=1.05em,font=\footnotesize](mk){#1};}

\newcommand{\ToyLongTPSPercent}{0.106}
\newcommand{\ToyLongFlatPercent}{8.12}

\usepackage[nameinlink]{cleveref}
\makeatletter
\AtBeginDocument{\pretocmd{\hyper@linkfile}{%
  \ifx\Hy@href@page\@empty\else
    \ifnum\Hy@href@page=0\relax\let\Hy@href@page\@empty\fi
  \fi}{}{\PackageError{paper}{Cannot initialize cross-document links}{}}}
\makeatother
\makeatletter
\newcommand{\externallabels}[2]{{\makeatletter\def\XR@prefix{}\let\XR@bibcite\vadjust\def\XR@URL{#2}%
  \filename@parse{#1}\XR@next#1\relax\\}}
\makeatother
\DeclareRobustCommand{\SICref}[1]{SI, \Cref{#1}}
\DeclareRobustCommand{\SIcref}[1]{SI, \cref{#1}}

\DeclareRobustCommand{\SIfirstcref}[1]{see the Supplementary Information, SI, \cref{#1}}
\crefname{equation}{Eq.}{Eqs.}
\crefname{algorithm}{Algorithm}{Algorithms}
\crefname{figure}{Figure}{Figures}
\crefname{table}{Table}{Tables}
\crefname{section}{Section}{Sections}
\crefname{subsection}{Section}{Sections}
\crefname{subsubsection}{Section}{Sections}
\crefname{theorem}{Theorem}{Theorems}
\crefname{lemma}{Lemma}{Lemmas}
\crefname{proposition}{Proposition}{Propositions}
\crefname{corollary}{Corollary}{Corollaries}
\crefname{definition}{Definition}{Definitions}
\crefname{assumption}{Assumption}{Assumptions}
\crefname{remark}{Remark}{Remarks}
\Crefname{equation}{Eq.}{Eqs.}
\crefname{appendix}{Appendix}{Appendices}
\crefname{subappendix}{Appendix}{Appendices}
\crefname{subsubappendix}{Appendix}{Appendices}
\Crefname{appendix}{Appendix}{Appendices}
\Crefname{subappendix}{Appendix}{Appendices}
\Crefname{subsubappendix}{Appendix}{Appendices}

\newcounter{resultheading}
\newcommand{\ResultHeadingName}{}

\newcommand{\ResultParagraph}[1]{%
  \renewcommand{\ResultHeadingName}{#1}%
  \refstepcounter{resultheading}%
  \paragraph{#1.}%
}
\crefformat{resultheading}{#2``#1''#3}
\Crefformat{resultheading}{#2``#1''#3}
\crefmultiformat{resultheading}{#2``#1''#3}{ and #2``#1''#3}{, #2``#1''#3}{, and #2``#1''#3}
\Crefmultiformat{resultheading}{#2``#1''#3}{ and #2``#1''#3}{, #2``#1''#3}{, and #2``#1''#3}

\AtBeginDvi{}
\hypersetup{colorlinks=true,linkcolor=opblue,citecolor=opblue,urlcolor=opblue,pdfauthor={Muhammad R. Hasyim},pdfsubject={Manuscript},pdftitle={What Do Co-Folding Models Learn? Insights from Enhanced Transition Path Sampling}}
\ifdefined\UseOneColumnSI
  \hypersetup{citecolor=MidnightBlue,linkcolor=BrickRed,filecolor=BrickRed,urlcolor=blue,pdfsubject={Supplementary Information}}
  
  \AtBeginDocument{%
    \renewcommand{\algorithmiccomment}[1]{%
      \hspace{1.2em plus 1fill}{\footnotesize\color{black!55}$\triangleright$~#1}%
    }%
  }
\fi

\externallabels{si-labels.tex}{si.pdf}
\icmltitlerunning{What Do Co-Folding Models Learn?}
\begin{document}
\raggedbottom
\twocolumn[
\icmltitle{What Do Co-Folding Models Learn? \\ Insights from Enhanced Transition Path Sampling}
\begin{icmlauthorlist}
\icmlauthor{Muhammad R.~Hasyim}{sccpc}
\end{icmlauthorlist}
\icmlaffiliation{sccpc}{The Simons Center for Computational Physical Chemistry, New York University, New York, NY 10003, USA}
\icmlcorrespondingauthor{Muhammad R.~Hasyim}{mh7373@nyu.edu}
\icmlkeywords{Transition path sampling, co-folding, enhanced sampling, OPES, replica exchange, diffusion models}
\vskip 0.3in
]
\printAffiliationsAndNotice{}
\begin{abstract}
Co-folding models are diffusion-based generative models that sample biomolecular complexes from distributions learned on a large corpus of experimental data.
Although these models are powerful for structure prediction, it remains unclear whether the conformational ensembles they learn resemble physically meaningful statistical ensembles.
To understand what these models learn, we introduce transition path sampling (TPS), which uses Markov chain Monte Carlo over denoising trajectories to sample the learned distribution. 
Because its Monte Carlo trajectories resemble biomolecular dynamics, TPS couples directly with enhanced-sampling techniques to recover the full conformational ensemble.
Applied to Boltz-2, enhanced TPS reveals physical conformational changes biased toward crystal-like states near the training data, indicating both learned physics and memorization.
These results suggest new ways to evaluate and improve co-folding models by jointly assessing structural recovery, proximity to training data, and conformational ensembles.

\end{abstract}
\section{Introduction}
\label{sec:intro}

\begin{figure*}[t]
\centering
\includegraphics[width=\textwidth]{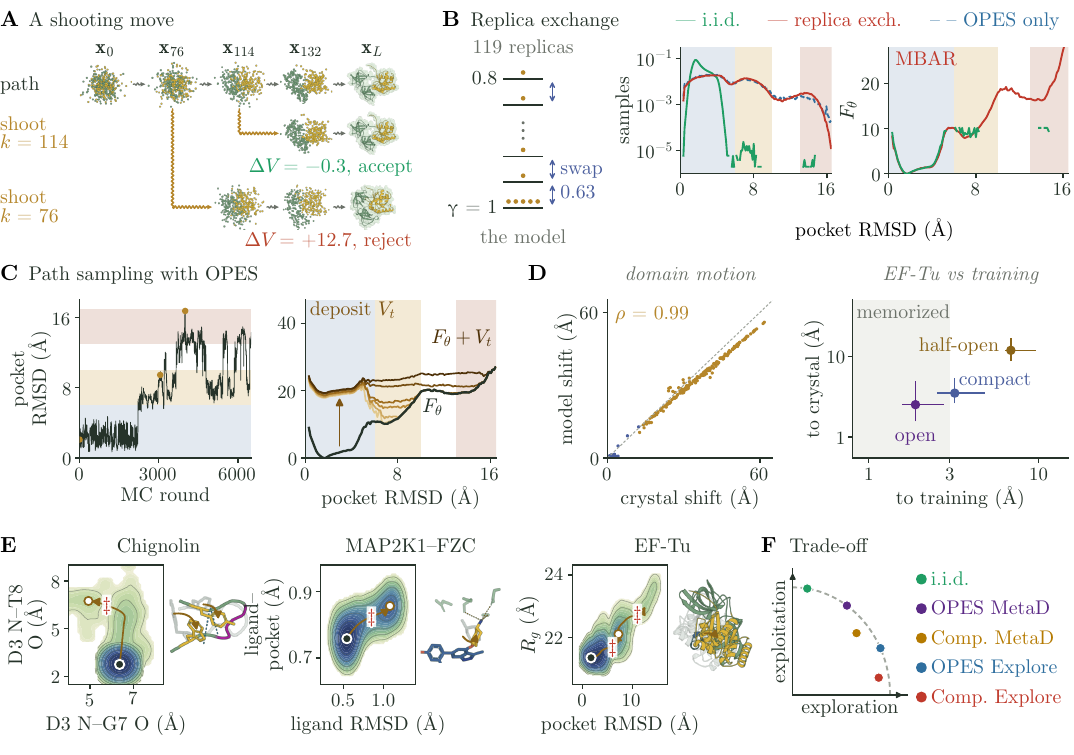}
\caption{Enhanced transition path sampling probes the conformational landscape of a co-folding model. 
(A)~A shooting move on EF-Tu (domain~I gold, domains II--III sage). Early shooting ($k=76$) changes the structure more than late shooting ($k=114$), with acceptance set by the bias potential difference $\Delta V$.
(B)~Replica exchange through $119$ tempered replicas ($\gamma=1$--$0.8$ setting effective 'temperature'), from which reweighting recovers $F_\theta$ beyond i.i.d. sampling. 
(C)~OPES on EF-Tu protein. The bias builds from $150$ Monte Carlo (MC) rounds (light) to $6{,}500$ (dark), and $F_\theta+V_t$ flattens as walkers reach the compact, half-open and open states (blue, sand, rose).
(D)~Model against experiment, where domain shifts follow crystal shifts (left) and only the half-open state lies far from training data (right). 
(E)~Landscapes and pathways in chignolin, MAP2K1-FZC (protein-ligand system) and EF-Tu protein, with reference states (dots), states (circles) and saddles ($\ddagger$).
(F)~Schematic trade-off (dashed) between exploration and exploitation for our proposed sampling methods. OPES is displayed in two modes, \textbf{MetaD} and \textbf{Explore}. Composite refers to OPES combined with score-tempered replica exchange.}
\label{fig:framework}
\end{figure*}

Biomolecules populate conformational states with probabilities set by equilibrium statistical mechanics.\citep{frauenfelder1991landscapes,henzlerwildman2007dynamic} 
From cryptic pockets to allosteric switches, the conformational ensemble governs recognition, regulation, and drug binding, making it crucial for understanding biological function.\citep{amaro2018ensemble}
Sampling these populations with molecular dynamics (MD) simulations is costly for large complexes, motivating deep generative models as an alternative.
Co-folding models like AlphaFold~3, OpenFold~3, Boltz, Chai, and Protenix form such a class of generative models, capable of sampling biomolecular complexes from sequence and ligand descriptions via reverse diffusion over atomic coordinates.\citep{abransom2024af3,passaro2025boltz2,chai2024,chen2025protenix,song2021score,karras2022edm} 
Although some co-folding models are now trained on conformational ensemble data,\citep{passaro2025boltz2} evaluations still emphasize structural accuracy,\citep{abransom2024af3,passaro2025boltz2} leaving the relationship between learned and physical ensembles poorly understood.

A growing set of evaluations now addresses this gap by probing the physics learned by co-folding models.
For example, a recent evaluation demonstrated that co-folding models can sample ligand-dependent cryptic-pocket openings.\citep{lazou2026cryptic} 
In general, however, successful predictions rely on similarity to the training set. 
As such, models can retain memorized poses after disruptive pocket mutations, placing allosteric ligands in orthosteric pockets,\citep{skrinjar2026memorization,masters2026generalization,nittinger2025cofolding} leave kinases and GPCRs in apo-like or inactive-like states despite accurate ligand placement,\citep{sun2026kinconfbench,obendorf2026decoupling} and maintain a single conformational state in the presence of new small molecules.\citep{ye2026multistate}
These failures matter because structural errors propagate downstream in drug discovery tasks, where misplaced ligands or expanded kinase pockets\citep{zhang2026gpcr,thaler2025boltzabfe} may corrupt the free-energy calculations used in lead optimization and candidate filtering.\citep{passaro2025boltz2}

Biophysical interactions dictate the conformational ensemble, so assessing whether co-folding models learn physics requires sampling the full ensemble of generated structures, including rare states.
To this end, several methods broaden sampling by perturbing the multiple sequence alignment (MSA) or internal representations,\citep{wayment2024afcluster,lee2026confornets} or by adding repulsion and walk-jump moves,\citep{ohnuki2025af3red,daigavane2026boltzjump} but these changes alter the sampled distribution without the reweighting factors needed to recover the original distribution.\citep{wayment2024afcluster}
Reweighted particle methods, as in ConforMix for Boltz-1 and enhanced diffusion sampling for BioEmu, can recover the original distribution \citep{richman2025conformix,xie2026enhanced} but rely on approximate steering and importance weights that degenerate when biased and target ensembles overlap poorly.
Therefore, sampling rare states in a co-folding model while accurately estimating their probabilities remains an open problem.

In this work, we introduce and apply transition path sampling (TPS)\citep{dellago1998transition} to co-folding models, an approach recently extended to language models.\citep{dorman2026rare}
TPS is a Metropolis--Hastings algorithm that recasts the diffusion process in co-folding models as a Markov chain over denoising trajectories, where detailed balance guarantees exact sampling of the learned distribution (Figure~\ref{fig:framework}A).
Successive proposal moves, also known as shooting moves, yield a Monte Carlo trajectory that resembles biomolecular dynamics, allowing TPS to inherit the rare-event machinery of molecular simulation.
We implement \textit{enhanced transition path sampling} with on-the-fly probability enhanced sampling (OPES)\citep{invernizzi2020opes,invernizzi2022opesexplore} and score-tempered replica exchange,\citep{sugita1999replica,fukunishi2002hamiltonian} the latter inspired by noise-level replica exchange in diffusion models.\citep{he2025crepe,omidi2026metalica} 
We can then reweight samples to estimate rare-state probabilities from the original model (Figure~\ref{fig:framework}B,C).
By analyzing the conformational ensemble of Boltz-2\citep{passaro2025boltz2} alongside experiment and training-set proximity,\citep{buttenschoen2024posebusters,skrinjar2026memorization} we find that the model captures physically reasonable trends while still over-preferring the native structure (Figure~\ref{fig:framework}D,E).
These estimates are subject to an exploration--exploitation trade-off (Figure~\ref{fig:framework}F).

In what follows, \Cref{sec:theory} develops the theory of enhanced TPS, from constructing the Markov Chain Monte Carlo to statistical reweighting. \Cref{sec:results} then applies the method to Boltz-2, first discussing the exploration-exploitation trade-offs, followed by case studies of several biological targets, ranging from a fast-folding protein to protein-ligand-nucleotide complexes.  

\section{Enhanced Transition Path Sampling}
\label{sec:theory}

Co-folding models generate structures by turning a noise vector into all-atom coordinates of a biomolecular complex.\citep{song2021score,karras2022edm}
Starting from random noise $\bm x_0$ under fixed conditioning $\bm c$, sampling follows from a stochastic differential equation (SDE) discretized into $L$ steps:
\begin{equation}
 \bm x_{j+1}=\mathcal T_j(\bm x_j,\bm\epsilon_j;\bm c),
 \qquad \bm\epsilon_j\sim\mathcal N(0,\nu_jI) \,,
 \label{eq:main-generator}
\end{equation}
where $\mathcal T_j$ is the update mapping and $\nu_j$ the noise variance.
For the Boltz-2 sampler,\citep{abransom2024af3} this mapping takes the form
\begin{equation}
\begin{aligned}
 \hat t_j&=\sigma_j(1+\chi_j),\\
 \hat{\bm x}_j&=(\bm x_j-\bar{\bm x}_j)\bm R_j+\bm t_j+\bm\epsilon_j,\\
 \bm u_j&=\bigl[D_\theta(\hat{\bm x}_j,\hat t_j;\bm c)-\hat{\bm x}_j\bigr]/\hat t_j^{\,2},\\
 \bm x_{j+1}&=\hat{\bm x}_j+\eta\,\hat t_j(\hat t_j-\sigma_{j+1})\,\bm u_j,\\
\end{aligned}
 \label{eq:main-step}
\end{equation}
where each update applies a random rigid motion $(\bm R_j,\bm t_j)$ about the centroid $\bar{\bm x}_j$, injects noise $\bm\epsilon_j$ to raise the noise level from $\sigma_j$ to $\hat t_j$ with the injection factor $\chi_j$, and takes an Euler step along $\bm u_j$, scaled by $\eta$ (\SIfirstcref{eq:af3-step}).
The learned component is the denoiser $D_\theta$, represented via a transformer architecture and trained using a score-matching objective.\citep{abransom2024af3,karras2022edm}

Applying \cref{eq:main-step} over $L$ steps transforms the initial noise into a structure $\bm x_L$ drawn from the learned conditional distribution $p_\theta(\bm x_L\,|\,\bm c)$---a process known as \textit{denoising}.
Repeated application of the denoising process then yields an ensemble of structures, which we analyze through collective variables (CVs), i.e., functions $s(\bm x)$ that distinguish conformational states like the root-mean-square deviation (RMSD) from the crystallographic pose.
In the CV space, the samples populate according to the marginal distribution 
\begin{equation}
 p_\theta(s)=\int\delta\bigl(s-s(\bm x_L)\bigr)\,p_\theta(\bm x_L\,|\,\bm c)\,d\bm x_L \,,
 \label{eq:main-cv-marginal}
\end{equation}
where we omit the dependency on $\bm c$ for brevity. The negative logarithm $F_\theta(s)=-\ln p_\theta(s)$ is the model landscape, which maps the conformational basins of a biomolecule. 

Generating structures in co-folding models proceeds as i.i.d. sampling, thus avoiding sample correlations that plague MD simulations. 
However, reaching a rare-event region $A$, populated with probability $p_A$, requires $1/p_A$ independent draws that diverges as $p_A\to0$.
MD simulations address this problem by biasing dynamics along CVs, an approach that recent work has carried over to the denoising process in co-folding models.\citep{richman2025conformix,xie2026enhanced} 
Such adoption of enhanced sampling relies on sequential Monte Carlo, whose finite-particle approximations generally do not yield exact target-distributed samples.\citep{delmoral2006sequential}
As shown in the next section, we instead adopt transition path sampling (TPS)\citep{dellago1998transition,dellago2002tps} as a framework for sampling $p_\theta( \bm x_L \, | \, \bm c)$ exactly. 

\begin{figure*}[!tp]
\centering
\includegraphics[width=\textwidth]{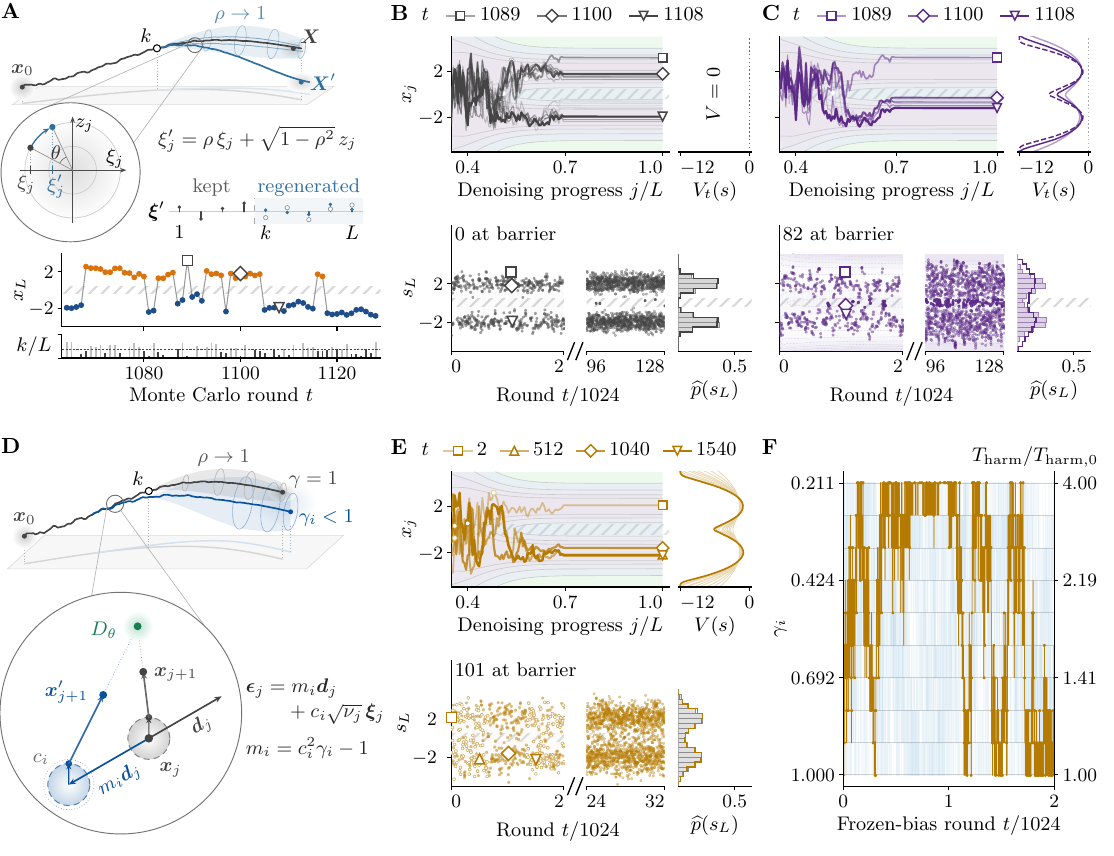}
\caption{Enhanced transition path sampling (TPS) using the analytic double-well diffusion process. 
(A) The shooting move regenerates the path beyond $k$ within a tube that narrows as $\rho\to1$. The inset shows one step whose input rotates toward a fresh draw by $\theta=\arccos\rho$; below, the terminal state $x_L$ and shooting index $k/L$ over Monte Carlo rounds. 
(B,C) (top) Denoising trajectories sampled by TPS and TPS--OPES with the bias potential $V_t$ over round $t$; (bottom) samples of $x_L$ over the first and last $2{,}048$ rounds with their histograms overlaid. $V_t$ in OPES converges to the tempered double-well, enabling samples in the transition-state region.
(D) Score tempering modifies path generation by shifting the mean and covariance of the noise inputs. The inset shows one tempered step (blue) shifted by $m_i\bm d_j$ against the carried score $\bm d_j$ and scaled by $c_i$. 
(E) (top) Denoising trajectories sampled by TPS with the bias potentials $V_t$ from tempered replicas overlaid in gold. 
(bottom) samples of $x_L$ from the base replicas, as in (B,C). Replica exchange covers the transition state with no bias potential.
(F) Walker assignments to replicas $\gamma_i$ over frozen-bias rounds. See \SICref{alg:tps,alg:tps-opes,alg:score-temper,alg:replica-exchange}.}
\label{fig:overview}
\end{figure*}

\subsection{Transition path sampling of diffusion models}
\label{sec:prelim}

TPS samples the trajectories of a co-folding model from the initial coordinates $\bm x_0$ to all possible output structures.
The algorithm achieves this via shooting moves, which perturb the sampler inputs $\bm\Omega$, comprising $\bm x_0$ and stochastic-step noises $\bm \epsilon_j$, i.e., $\bm\Omega=(\bm x_0,\{\bm\epsilon_j\}_{j\in\mathcal K})$. 
These inputs have the product Gaussian reference distribution
\begin{equation}
 \mu_0(d\bm\Omega)=\rho_0(d\bm x_0)\prod_{j\in\mathcal K}\mathcal N(d\bm\epsilon_j;0,\nu_jI).
 \label{eq:main-seed}
\end{equation}
For given inputs, iterating \cref{eq:main-generator} for $L$ steps deterministically generates a denoising trajectory, defining the map $\bm X=(\bm x_0,\ldots,\bm x_L)=\mathcal F_{\bm c}(\bm\Omega)$.
The corresponding push-forward distribution $\Pi_{\bm c}=\mathcal F_{\bm c\#}\mu_0$ is the path distribution induced by \cref{eq:main-seed} with terminal marginal $p_\theta(\bm x_L\,|\,\bm c)$.\citep{dellago2002tps}
To resolve rare states along the CVs $s(\bm x)$, we tilt the ensemble at $s_L=s(\bm x_L)$ by a bias potential $V(s_L)$:
\begin{equation}
 \Pi_V=Z_V^{-1}e^{-V(s_L)}\Pi_{\bm c},
 \qquad
 \widetilde\Pi_V=Z_V^{-1}e^{-V(s_L)}\mu_0,
 \label{eq:main-target}
\end{equation}
where $Z_V=\int e^{-V(s_L)}\mu_0(d\bm\Omega)$ and reweighting by $e^{V(s_L)}$ recovers expectations under $p_\theta$.
Tilts that depend only on the terminal structure have precedents in umbrella sampling of path ensembles, where windows restrict the order parameter at the end point of each trajectory,\citep{dellago2002tps} and in reward-guided diffusion sampling, where the target is $p_\theta(\bm x_L)\,e^{\lambda r(\bm x_L)}$ for a reward $r$ on the generated sample.\citep{singhal2025steering}

As a Metropolis-Hastings algorithm, TPS relies on a well-chosen proposal move. 
Following \citet{cotter2013}, we adopt the so-called shooting move that selects an index $k$ according to a distribution $\pi_k$, retains the path segment $\bm x_{\le k}$, and regenerates the remaining segment by perturbing the standardized noises $\bm\xi_j=\bm\epsilon_j/\sqrt{\nu_j}$ for $j\ge k$ as follows:
\begin{equation}
 \bm\xi_j'=\rho\bm\xi_j+\sqrt{1-\rho^2}\bm z_j\;\;(j\ge k),
 \label{eq:main-pcn}
\end{equation}
where $\rho\in(0,1)$ is a correlation parameter and $\bm z_j\sim\mathcal N(0,I)$ are independent Gaussian inputs (\Cref{fig:overview}A).
Geometrically, \cref{eq:main-pcn} rotates $(\bm\xi_j,\bm z_j)$ by $\theta=\arccos\rho$, so for $\rho$ near one the new path stays close to the current path (\Cref{fig:overview}A).

To construct the acceptance criterion, we write the proposal kernel, i.e., the distribution of the proposed inputs $\bm\Omega'$ given the current inputs $\bm\Omega$, as
\begin{equation}
\begin{aligned}
 Q(\bm\Omega,d\bm\Omega')=\sum_k\pi_k
 &\prod_{j<k}\delta_{\bm\xi_j}(d\bm\xi_j')\\
 \times&\prod_{j\ge k}\mathcal N\bigl(d\bm\xi_j';\rho\bm\xi_j,(1-\rho^2)I\bigr).
\end{aligned}
 \label{eq:main-proposal}
\end{equation}
Under $\mu_0$, each pair $(\bm\xi_j,\bm\xi_j')$ ($j\ge k$) is a symmetric bivariate Gaussian with unit variances and covariance $\rho I$, making $Q$ reversible with respect to $\mu_0$, i.e., $\mu_0(d\bm\Omega)Q(\bm\Omega,d\bm\Omega')=\mu_0(d\bm\Omega')Q(\bm\Omega',d\bm\Omega)$.\citep{cotter2013}
The reversibility of $Q$ with respect to
$\mu_0$ simplifies the Metropolis--Hastings ratio for
$\widetilde\Pi_V$, a Radon--Nikodym derivative on the product space, to a Boltzmann factor, i.e.,
\begin{equation}
\begin{aligned}
 \frac{\widetilde\Pi_V(d\bm\Omega')\,Q(\bm\Omega',d\bm\Omega)}{\widetilde\Pi_V(d\bm\Omega)\,Q(\bm\Omega,d\bm\Omega')}
 &=e^{V(s_L)-V(s_L')}\,\frac{\mu_0(d\bm\Omega')\,Q(\bm\Omega',d\bm\Omega)}{\mu_0(d\bm\Omega)\,Q(\bm\Omega,d\bm\Omega')}\\
 &=e^{V(s_L)-V(s_L')} \,.
\end{aligned}
 \label{eq:main-mh-ratio}
\end{equation}
Since the reference measure and the proposal cancel, only the terminal bias remains in the acceptance probability:
\begin{equation}
 a(\bm\Omega,\bm\Omega')
 =\min\bigl\{1,e^{V(s_L)-V(s_L')}\bigr\}.
 \label{eq:main-accept}
\end{equation}
Thus, accepting the proposed move with probability \cref{eq:main-accept} ensures detailed balance with respect to $\widetilde\Pi_V$.

We demonstrate TPS on a one-dimensional diffusion model in \Cref{fig:overview}A(bottom) and B, for which transport from a Gaussian to a symmetric mixture of two Gaussians is analytically solvable.
The score has a single attractor early in denoising, which splits into two basins once the noise falls below the component separation (\Cref{fig:overview}B).
A shoot before this split sends the new trajectory into either basin.
As shown in \Cref{fig:overview}A (bottom), this effect appears in the sampled terminal state $x_L$, whose apparent dynamics execute rare hops between the two basins. 

\subsection{Enhanced sampling with OPES}
\label{sec:opes}
To promote exploration, we may choose a bias potential $V(s)$ that targets a distribution $p^{\rm tg}(s)$ that samples the CV space more favorably.
To obtain $p^{\rm tg}(s)$ from $p(s)\equiv p_\theta(s)$ in \cref{eq:main-cv-marginal}, we can derive the ideal bias potential following the identity $p_V(s)\propto p(s)e^{-V(s)}$, which implies\citep{torrie1977umbrella}
\begin{equation}
 \qquad V(s)=\ln\frac{p(s)}{p^{\rm tg}(s)} \,,
 \label{eq:main-opes-target}
\end{equation}
up to an additive constant. A tunable choice is the well-tempered marginal $p^{\rm tg}(s)\propto p(s)^{1/\gopes}$, with $\gopes>1$, which leads to the following ideal bias \citep{invernizzi2020opes}
\begin{equation}
 \qquad V(s)=-(1-\gopes^{-1})F(s) \,.
 \label{eq:main-opes-wt}
\end{equation}
\Cref{eq:main-opes-target} implies that the ideal bias reduces the barriers inherent to $F(s)$ by a factor $\gopes$ and approaches uniform sampling as $\gopes\to\infty$.
However, \Cref{eq:main-opes-target} reveals a central challenge in enhanced sampling as the ideal bias requires estimating the initially unknown $F(s)$. 
On-the-fly probability enhanced sampling (OPES)\citep{invernizzi2020opes,invernizzi2020unified} addresses this circular dependence by estimating $F(s)$ during sampling and updating the bias after each Monte Carlo round.

In OPES \textbf{MetaD} mode, \citep{invernizzi2020opes} adaptive, weighted kernel density estimation reconstructs $p(s)$, denoted by $\hat p(s)$.
OPES alternates rounds of TPS, each comprising multiple Monte Carlo steps at fixed bias.
At the end of each round, OPES adds a Gaussian kernel $G(s;\zeta_a,\Sigma_a)$ centered at the current terminal CV $\zeta_a$ to the current estimate of $\hat p_t(s)$ at round $t$, leading to an expression
\begin{equation}
 \hat p_t(s)=\frac{1}{\mathcal N_t}\sum_{a=1}^{n_K}h_aG(s;\zeta_a,\Sigma_a),
 \quad \mathcal N_t=\sum_{a=1}^{n_K}h_a,
 \label{eq:main-opes-kde}
\end{equation}
where $h_a=e^{V_{t_a}(\zeta_a)}$ reweights the density estimate so that we appropriately construct the bias potential. 
\Cref{eq:main-opes-kde} then defines the updated bias,
\begin{equation}
 V_t(s)=\kappa\ln\Bigl[\frac{\hat p_t(s)}{Z_t}+\varepsilon_0\Bigr],
 \label{eq:main-opes}
\end{equation}
where $\kappa=1-\gopes^{-1}$ and $Z_t$ normalizes the estimate over the explored CV region.
The floor $\varepsilon_0=e^{-\Delta E/\kappa}$ limits the bias to $-\Delta E$ in the limit of vanishing estimated density.
Alternatively, \textbf{Explore} mode\citep{invernizzi2022opesexplore} uses equal weights $h_a=1$ to estimate the sampled marginal, which approaches $p^{\rm tg}(s)\propto p(s)^{1/\gopes}$, and sets $\kappa=\gopes-1$ to favor exploration over rapid convergence to a quasi-static bias (\Cref{sec:boltz-overview}).

The double well (\Cref{fig:overview}C) illustrates how the adaptive bias changes sampling.
OPES MetaD on $s=x_L$, with $\gopes=\infty$ ($\kappa=1$) and $\beta\Delta E=12$, starts from the same paths and proposal draws as TPS.
It initially behaves like TPS because the bias acts only through the acceptance test.
As kernels accumulate, $V_t$ rises in the wells relative to the barrier (side curves), promoting barrier sampling.
Over the final set of rounds, $\ToyLongFlatPercent\%$ of samples occupy the transition state ($|x_L|<0.4$), compared with $\ToyLongTPSPercent\%$ for TPS.

\subsection{Score Tempering and Replica Exchange}
\label{sec:tempering}

When the CVs omit slow orthogonal coordinates, OPES alone may leave simulations behind hidden barriers.
To overcome such barriers, we may also use replica exchange,\citep{sugita1999replica,fukunishi2002hamiltonian} a technique from MD simulations in which several replicas of the system, simulated at a ladder of temperatures, exchange configurations to promote sampling at colder temperatures.
Here, a configuration is an entire denoising path, i.e., its sampler inputs $\bm\Omega$.
To emulate a change in temperature, each replica instead draws its noises from a tempered distribution, while the denoiser and the update of \cref{eq:main-step} remain those of the base model.
Replica $i$ shifts the mean of each noise along the score of the previous step with a score scale $\gamma_i<1$, and it may also widen the noise with a noise scale $c_i$.
For a reverse-SDE step with mean $\bm x_j+\nu_j\bm u_j$, this shift is equivalent to multiplying the score by $\gamma_i$, and hence we refer to this technique as \emph{score tempering}.\citep{skreta2025feynman}

Score tempering may be defined by defining the modified noise distribution. Given standardized inputs $\bm\xi_j\sim\mathcal N(0,I)$, replica $i$ under score tempering samples its noise as
\begin{equation}
 \begin{aligned}
 \bm\epsilon_j&=m_i\bm d_j+c_i\sqrt{\nu_j}\bm\xi_j,
 \qquad m_i=c_i^2\gamma_i-1,\\
 \bm d_j&=\nu_j\bm u_{j-1}\bm R_j,\qquad \bm d_0=0,
 \end{aligned}
 \label{eq:main-tempering}
\end{equation}
with score scale $\gamma_i$ and noise scale $c_i$ ($\gamma_0=c_0=1$).
\Cref{fig:overview}D visualizes score tempering in replica $i$.
Given the prefix, $\bm\epsilon_j\sim\mathcal N(m_i\bm d_j,c_i^2\nu_jI)$, and these conditionals define the tempered input distribution $\mu^{(i)}$. 
To tune the temperature ladder, set by the factors $\gamma_i$, we show in \SICref{methods:sec:tempering} that the spacing is determined by the Kullback--Leibler divergence between neighboring input distributions.

Since a replica generates trajectories with different noise inputs, we need to track the path weight $\ell_i(\bm\Omega)=\ln\frac{d\mu_0}{d\mu^{(i)}}(\bm\Omega)$, the logarithm of the Radon--Nikodym derivative of the base input distribution ($i=0$) with respect to that of replica $i$. 
Intuitively, $\ell_i(\bm \Omega)$ compares the probability that the base replica produces paths with sampler inputs $\bm\Omega$ with the probability that replica $i$ produces the same paths.
Since every $\bm\epsilon_j$ is Gaussian, the log-ratio decomposes into
\begin{equation}
 \ell_i(\bm\Omega)=\sum_{j\in\mathcal K}
 \frac{-m_i\bm\epsilon_j\cdot\bm d_j+\tfrac12m_i^2\|\bm d_j\|^2}{\nu_j},
 \label{eq:main-path-weight}
\end{equation}
for score-only ladders ($c_i=1$, $m_i=\gamma_i-1$), and $\mu^{(i)}=e^{-\ell_i}\mu_0$. The path weight thus plays the role of an energy, and the input distribution of replica $i$ is the base distribution reweighted by a factor $e^{-\ell_i}$.
Each replica also tilts its distribution by its own OPES bias $V_i$ of \cref{eq:main-opes}, so the path probability of replica $i$ is
\begin{equation}
 \widetilde\Pi_i(d\bm\Omega)\propto e^{-V_i(s_L)}\mu^{(i)}(d\bm\Omega)
 =e^{-U_i}\mu_0(d\bm\Omega) \,,
 \label{eq:main-replica-target}
\end{equation}
where $U_i=V_i+\ell_i$. The shooting move of \cref{eq:main-pcn} on $\bm\xi_j$ is reversible with respect to $\mu^{(i)}$, so \cref{eq:main-accept} preserves $\widetilde\Pi_i$.

After every round, neighboring replicas $i$ and $j=i+1$ attempt to exchange sampler inputs in even--odd sweeps (\SICref{alg:replica-exchange}); detailed balance for $\widetilde\Pi_i\otimes\widetilde\Pi_j$ requires
\begin{equation}
 \frac{\widetilde\Pi_i(d\bm\Omega_j)\,\widetilde\Pi_j(d\bm\Omega_i)}
      {\widetilde\Pi_i(d\bm\Omega_i)\,\widetilde\Pi_j(d\bm\Omega_j)}
 =e^{\Delta V_{ij}}\,
  \frac{\mu^{(i)}(d\bm\Omega_j)\,\mu^{(j)}(d\bm\Omega_i)}
       {\mu^{(i)}(d\bm\Omega_i)\,\mu^{(j)}(d\bm\Omega_j)},
 \label{eq:main-swap-ratio}
\end{equation}
where $\Delta g_{ij}=g_i(\bm\Omega_i)+g_j(\bm\Omega_j)-g_i(\bm\Omega_j)-g_j(\bm\Omega_i)$ for any replica-indexed number $g$.
Unlike in MD, the $\mu$ factor does not cancel, because the replicas draw their noises from different distributions.
The exchange is then equivalent to a Hamiltonian replica-exchange move,\citep{fukunishi2002hamiltonian} with $U_i$ playing the role of the energy function,
\begin{equation}
 a_{ij}=\min\{1,e^{\Delta U_{ij}}\},
 \qquad \Delta U_{ij}=\Delta V_{ij}+\Delta\ell_{ij},
 \label{eq:main-swap}
\end{equation}
which preserves the product of replica targets (\SICref{app:stationarity}).

In addition to exchanging paths, another way to promote sampling from hotter replicas is to construct a new biasing potential with their computed path weights $\ell_i$.
Following OneOPES,\citep{rizzi2023oneopes} we note that the path weights may carry information omitted by the CVs, as they depend on every noise generating the entire structure.
This intuition motivates the OPES-Expanded \textbf{multithermal} bias potential\citep{invernizzi2020unified} that can be added to select replicas $i\ge i_\star$ ($i_\star=4$),
\begin{equation}
 B_i(\bm\Omega)=V_i(s_L)-\mathbf 1_{i\ge i_\star}\ln\frac1{N_\lambda}\sum_{q=1}^{N_\lambda}e^{\lambda_q\ell_i(\bm\Omega)+\Delta f_q},
 \label{eq:main-bias-total}
\end{equation}
which replaces $V_i$ above ($\Delta B_{ij}$ and $U_i=B_i+\ell_i$).
The multithermal term changes the target of replica $i$ into a mixture of the distributions $\propto\mu^{(i)}e^{\lambda_q\ell_i}$ with $\lambda_q\in[\lambda_{\max},0]$.
In effect, it lets a replica bias sampling toward ``hotter'' regions without waiting for accepted swaps. 
Its offsets $\Delta f_q=-\ln Z_q$, with $Z_q$ the mean of $e^{\lambda_q\ell_i}$ excluding the multithermal term, are estimated as a bias-reweighted running average; see \SICref{methods:sec:tempering} for details.

In the double well (\Cref{fig:overview}D), the hottest replica, $\gamma=0.8$, widens the wells by $12\%$ and raises the barrier population $2.5$-fold (\SICref{methods:toy-validation}); under OPES with exchange, crossings on hot replicas reach the base replica by accepted swaps (\Cref{fig:overview}E,F). 

\subsection{Landscape Estimation}
\label{sec:landscape-reconstruction}

Samples from different replicas come from different statistical ensembles, which we pool to calculate the landscape via reweighting. 
While standard reweighting treats the bias as fixed, adaptive biasing such as OPES converges quasi-statically.\citep{invernizzi2020opes} 
To adapt reweighting techniques to OPES, we treat  each epoch, i.e., a collection of rounds, in a single OPES run as samples from a static statistical ensemble and perform time-dependent reweighting.\citep{giberti2020iterative} 
Here, we take the simplest version of this pooling strategy with MBAR.\citep{shirts2008mbar} For $S$ states $a$ (replicas or bias epochs) with potentials $U_a=B_a+\ell_a$ and $N_a$ samples $\bm\Omega_n\sim\widetilde\Pi_a\propto e^{-U_a}\mu_0$, MBAR solves the coupled equations
\begin{gather}
 f_a=-\ln\sum_{n=1}^{N}\frac{e^{-U_a(\bm\Omega_n)}}{\sum_{b=1}^{S}N_b\,e^{f_b-U_b(\bm\Omega_n)}}, \\
 W_n=\Bigl[\sum_{b=1}^{S}N_b\,e^{f_b-U_b(\bm\Omega_n)}\Bigr]^{-1},
 \label{eq:main-mbar}
\end{gather}
where $a=1,\dots,S$, for the free energies $f_a$ (up to a constant) and the weights $W_n$ that carry all $N=\sum_aN_a$ samples to $\mu_0$ ($U=0$). 
Smoothing the terminal CVs with kernel $K_h$ yields the reconstructed density $\hat p(s)=\sum_n W_nK_h(s-s_n)/\sum_nW_n$ and the landscape $\widehat F(s)=-\ln\hat p(s)+C$.

\begin{figure*}[!t]
\centering
\includegraphics[width=0.975\textwidth]{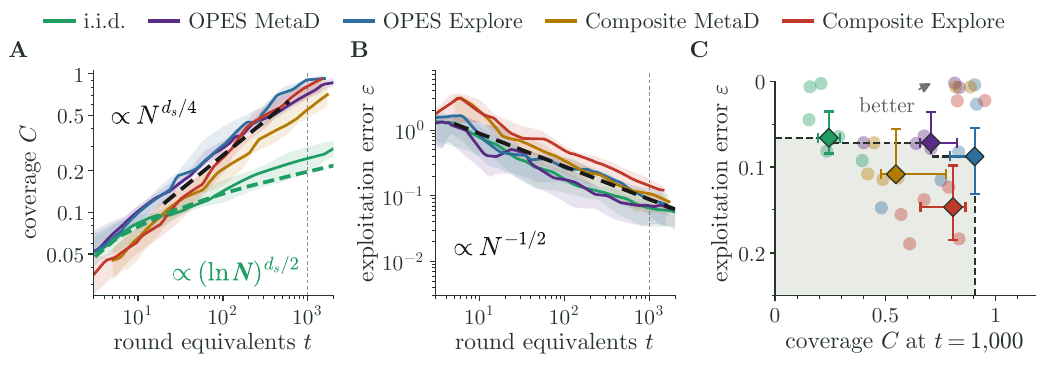}
\caption{Exploration and exploitation in enhanced TPS across various targets. 
(A)~Coverage $C$ of the supported CV space and (B)~exploitation error $\varepsilon$ against equivalent sampling rounds $t$. 
Each curve and shaded area displays the medians and interquartile ranges over targets. Dotted line denotes the $t=1000$ line. 
The dashed line represents the parameter-free scaling law of \cref{eq:main-coverage-law} for independent generation (green) and, with fitted prefactors, the random-walk scaling $C\propto t^{d_s/4}$ of \SIcref{eq:main-exponent-kinetic} and the convergence rate $\varepsilon\propto t^{-1/2}$ (black). (C)~Coverage against error at $t=1{,}000$. Diamond markers and error bars represent median and interquartile bars, with individual target performance marked in faded circles. Pareto front is drawn in dashed lines. See \SICref{si:ana-scores,fig:coverage-law,fig:height-diffusion,fig:error-walk}.}
\label{fig:tradeoff}
\end{figure*}

\section{Results}
\label{sec:results}

With enhanced TPS, a co-folding model runs as a Monte Carlo simulation that maps its learned landscape along chosen CVs.
In this section, we apply the algorithm to Boltz-2\citep{passaro2025boltz2} on short peptides, single proteins and biomolecular complexes,\citep{kjeldgaard1993eftu,berchtold1993eftu,xu1999csrc,perry2009p38,takano2023mek,skrinjar2026memorization,honda2004chignolin,kuhrova2012chignolin} first weighing enhanced sampling against i.i.d. generation (\Cref{sec:boltz-overview}).
We then examine the learned landscape of each system, from single funnels (\Cref{sec:funnels}) and two-well systems (\Cref{sec:double-well}) to the translation factor EF-Tu in its free and complexed states (\Cref{sec:eftu}).

Each case study is compared with experimental structures from the PDB\citep{honda2004chignolin,xu1999csrc,cowanjacob2005csrc,takano2023mek,wang1998p38,perry2009p38,kjeldgaard1993eftu,berchtold1993eftu,polekhina1996eftu,nissen1995ternary,vogeley2001aurodox,skrinjar2026memorization} and with free energies from experiments and MD simulations.\citep{kuhrova2012chignolin,yang2010hck,huang2020hck,johansen2018eftu}
Where such references are lacking, we compare the model with a force field along pathways between its basins.
Each pathway is a string\citep{e2002string} that slices the model landscape $\hat{F}(s)$ in one dimension, and we evaluate its energetics by adiabatic mapping,\citep{lonsdale2012practical} also known as the drag method. In this calculation, a structure is moved along the string in small steps using a force field (e.g., ff14SB/GAFF or AMARO-v1), and its energy is minimized with the CVs fixed at each image (\SICref{app:string-drag}).
Since the outcome can depend on the direction of the mapping,
we choose to map each string forward from the deeper of its two end states. 
Disagreement between the two profiles flags possible hallucinations of states.

Finally, to look beyond pose accuracy, we use an extension of SuCOS, the training-similarity score of the Runs N' Poses benchmark, as a collective variable that can be biased using enhanced TPS (\Cref{sec:experiment}; \Cref{fig:framework}), opening evaluations of memorization in conformational ensembles.

\subsection{Exploration vs. exploitation}
\label{sec:boltz-overview}\label{sec:matrix}

In this section, we present results on the exploration-exploitation trade-offs between i.i.d. sampling and enhanced TPS.
We assess exploration by dividing the CV space into bins and defining the bounded domain $\mathcal S$ as the set of bins where the model landscape $F_\theta(s) < 25$. 
From here, we may define coverage $C(t)$ as the fraction of the support $\mathcal S$ visited at least once after $t$ rounds. 
With $n_b(t)$ its number of samples in bin $b$, we write the coverage as 
\begin{equation}
 C(t)=\frac{1}{|\mathcal S|}\sum_{b\in\mathcal S}\mathbf 1\bigl[n_b(t)\ge1\bigr],
 \label{eq:main-coverage}
\end{equation}
where $\mathbf 1[\cdot]$ is the indicator function. 
Expressing $t$ in terms of equivalent rounds of i.i.d. sampling (\SICref{app:sampling-metrics}), we obtain a curve such as 
\Cref{fig:tradeoff}A that shows how every enhanced-sampling method outpaces i.i.d. sampling with specific scaling exponents. 

To understand the scaling growths in $C(t)$, we may view these algorithms as performing a random walk in CV space  (\SICref{app:scaling-derivations}).
For i.i.d. sampling, with $F_b$ the value of $F_\theta$ in bin $b$, $p_b=e^{-F_b}$ its model probability, $N_w$ the number of walkers, each of which adds one sample per round, and $\nu(f)$ the fraction of the support with $F_b<f$, the expected coverage is exactly
\begin{equation}
\begin{aligned}
 C(t)&=\frac{1}{|\mathcal S|}\sum_{b}\bigl(1-e^{-N_wt\,p_b}\bigr) =\mathbb E_Z\bigl[\nu(\ln N_wt+Z)\bigr],
\end{aligned}
 \label{eq:main-coverage-law}
\end{equation}
where $Z$ is a standard Gumbel variable.
\Cref{eq:main-coverage-law} predicts the whole i.i.d. curve within $3\%$ with no free parameter (\SICref{fig:coverage-law,eq:coverage-law}). 
On a surface harmonic about its minimum $F_0$ in $d_s$-many CVs, its scaling exponent is $d\ln C/d\ln t\approx d_s/[2\ln(N_wt\,p_0)]$, where $p_0=e^{-F_0}$ is the probability of the most populated bin (\SIcref{eq:main-exponent}).
Inserting $d_s=2$ and the minima $F_0$ from all targets reproduces the fitted exponents in \Cref{fig:tradeoff}A, which span $0.15$--$0.21$.

\begin{figure*}[!t]
\captionsetup{hypcap=false,skip=4pt,font=footnotesize}
\centering
\includegraphics[width=0.9\textwidth]{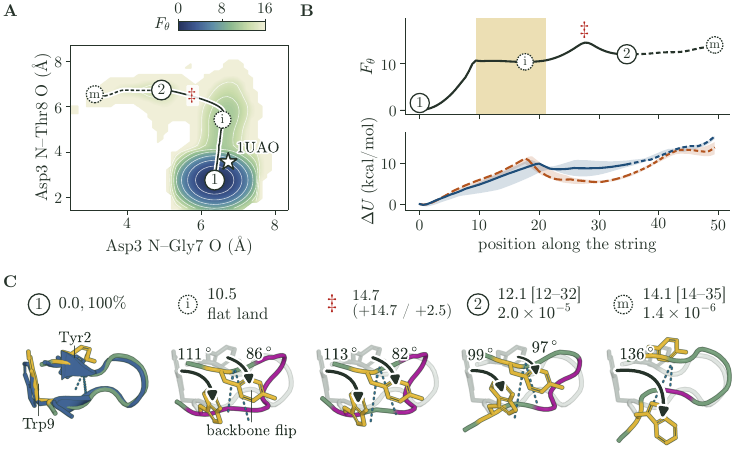}
\caption{Chignolin, a single funnel, pooled over five methods. (A)~The landscape $F_\theta$ over backbone distance CVs. 
(B)~The string profile above the energy $\Delta U$ from adiabatic mapping (ff14SB solid, AMARO dashed; energy profiles plots the median with the interquartile range; See \SICref{app:case-figure-conventions} for further details). 
(C)~The structures at the marked states, each over state~\stmark{1} in gray, with Tyr2 and Trp9 in gold, an arc for the swing of each ring about its strand, and in magenta the backbone residues whose $\phi$ or $\psi$ turns by more than $90^\circ$ from state~\stmark{1}; 1UAO slate. Star, 1UAO NMR ensemble; \stmark{m}, MD's misfolded hairpin on the dashed string.}
\label{fig:case-chignolin}
\end{figure*}

In contrast, enhanced sampling yields walkers with memory.
Under a flattening bias, the height $u=F_\theta-F_0$ of a walker above the minimum diffuses with a coefficient $D_u$ independent of height (\SICref{fig:height-diffusion}).
By the self-similarity of this diffusion, the coverage of a harmonic basin then grows as $C\propto(D_ut)^{d_s/4}$ (\SIcref{eq:main-exponent-kinetic}), and its exponent $d_s/4=1/2$ reproduces the $0.53$ fitted in \Cref{fig:tradeoff}A.

Exploitation is measured by the landscape error $\varepsilon(t)$, the RMSD between the surface $\hat F_t$ estimated after $t$ rounds and the reference $F_{\rm ref}$ (\SIcref{eq:fe-error}).
This error decays as a power of $t$ for every sampling method (\Cref{fig:tradeoff}B), which again follows from a random-walk picture.
Each free energy $F_b=-\ln\pi_b$ is estimated from bin counts, and its counting noise shrinks with the number of independent visits, spaced by the mean first-passage time $\mathbb E[\tau_b]$ of a stationary walk (\SIcref{eq:occupation-variance}).
Averaging over bins gives
\begin{equation}
    \varepsilon^2(t)\approx\frac{A_2}{N_wt}+\varepsilon_{\rm ref}^2,\quad A_2=\Bigl\langle2\,\mathbb E[\tau_b]+1-\frac{1}{\pi_b}\Bigr\rangle_\Omega \,,
    \label{eq:main-error-iid}
\end{equation}
where $\varepsilon_{\rm ref}$ is the error floor of the reference.
Independent samples give the binomial shot noise $A_2=\langle1/p_b\rangle_\Omega-1$, and the local exponent $d\ln\varepsilon/d\ln t=-\tfrac12A_2/(A_2+N_wt\,\varepsilon_{\rm ref}^2)$ converges to $-\tfrac12$ at the correct asymptotic regime, reproducing the fitted exponent.
Correlations between OPES samples lengthen the first-passage time by a memory time $T_{\rm mem}$, which raises the prefactor to $A_2\approx\langle1/\pi_b\rangle_\Omega-1+2T_{\rm mem}$ (\SIcref{eq:main-memory-time}) and slows convergence with the $-\tfrac12$ exponent.

Coverage and error at $t=1{,}000$ place the methods on a Pareto front (\Cref{fig:tradeoff}C) that runs from i.i.d. sampling through OPES MetaD to OPES Explore, trading a small rise in error for a large gain in coverage.
The score-tempering ladder of Composite MetaD and Composite Explore pays off only on targets where faster barrier crossing repays the cost of its replicas, and on the examples here that cost exceeds the gain, leaving both methods behind the front.

\subsection{Single Funnel Systems}
\label{sec:funnels}
With these tradeoffs established, we now examine the learned conformational ensembles, beginning with systems that display a single dominant state.

\paragraph{Chignolin.}

We begin with chignolin, a designed ten-residue $\beta$-hairpin whose folding equilibrium has been measured by circular dichroism and NMR.\citep{honda2004chignolin,kuhrova2012chignolin}
To sample its ensemble, we bias two backbone distances, Asp3 N--Thr8 O and Asp3 N--Gly7 O, which distinguish the folded hairpin from separated strands.\citep{satoh2006chignolin} 
As shown in \Cref{fig:case-chignolin}A, enhanced sampling with Boltz-2 reveals a single funnel around the folded hairpin (state~\stmark{1}), next to the 1UAO NMR ensemble (star).
However, the chance of leaving this basin is $0.0025\%$ (\SIcref{eq:chance}), whereas experiments place the unfolded state almost level with the folded one, at a $45\%$ chance.
Boltz-2 thus recovers the native structure but overestimates the stability of the folded state (\SICref{fig:evaluations}B). 

\begin{figure*}[!t]
\captionsetup{hypcap=false,skip=4pt,font=footnotesize}
\centering
\includegraphics[width=0.95\textwidth]{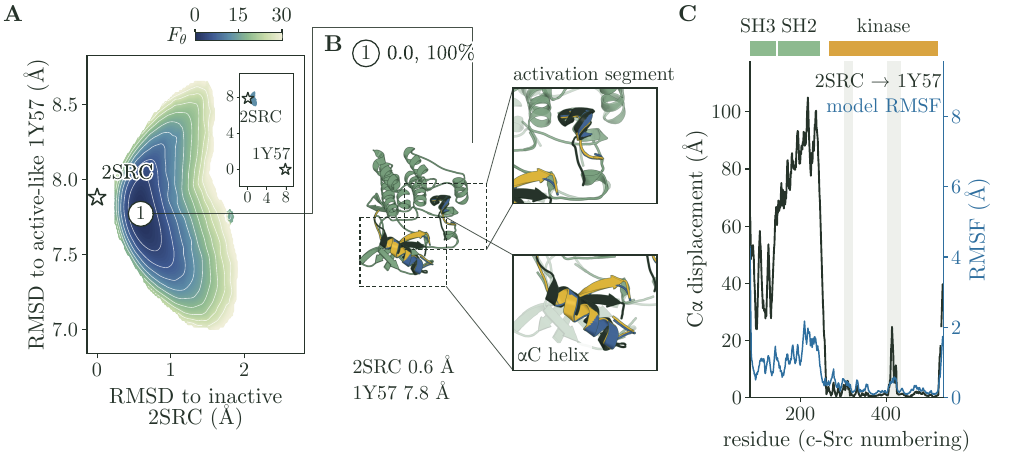}
\caption{Src kinase, a single funnel, on the RMSDs to inactive 2SRC and active-like 1Y57 (stars). 
(A)~The landscape with the surface extended to $F_\theta=30$. The inset shows the full plane spanning both active and inactive states, where the sampled region (blue) lies $7$--$9\,\text{\AA}$ from 1Y57. 
(B)~state~\stmark{1} on the whole kinase with 2SRC (slate) and 1Y57 (dark); the dashed boxes open into close-ups of the activation segment and the $\alpha$C helix. 
(C)~The C$\alpha$ displacement between the crystals against the model's RMSF (rank correlation $0.84$, against $0.56$ for distance from the aligned core), along the SH3, SH2 and kinase domains (ribbon), with shading on $\alpha$C and the activation segment.}
\label{fig:case-src}
\end{figure*}

Despite the dominance of the native structure, Boltz-2 also predicts a second basin on these CVs, state~\stmark{2} (\Cref{fig:case-chignolin}C).
This state has no correspondence in MD simulations or experiments, which instead identify a misfolded hairpin (\stmark{m}), with Tyr2 and Trp9 on opposite faces.\citep{kuhrova2012chignolin,maruyama2019chignolin}
Nevertheless, the string in \Cref{fig:case-chignolin}B connects the funnel with state~\stmark{2}.
The peptide first crosses a plateau along this string (\stmark{i}, shaded in the profile), with both hydrogen bonds broken but the aromatic packing intact.
A saddle ($\ddagger$) then separates this plateau from state~\stmark{2}, and state~\stmark{m} lies just beyond state~\stmark{2} on the continued string (dashed line).

Next, we perform adiabatic mapping along the string using the ff14SB and AMARO force fields (\SICref{app:string-drag}).
As shown in \Cref{fig:case-chignolin}B (bottom), the energy rises by $10$--$11$~kcal/mol upon breakage of the hydrogen bonds.
However, the flat land indicated in the model landscape is a barrier in the energy calculations, whereas the saddle is a shallow basin, before the energy climbs further toward states~\stmark{2} and \stmark{m}.
Toward the end of the string, the minimized peptide departs from the model structures (\SICref{fig:drag-structures}), and thus the force field does not reproduce state~\stmark{2} even structurally.
Overall, these results show that Boltz-2 predicts the native structure and, qualitatively, the physics of unfolding, but fails quantitatively.

\paragraph{Src kinase.}

Src is a multidomain tyrosine kinase with known inactive and active structures.\citep{xu1999csrc,cowanjacob2005csrc}
In the inactive form, its SH2 and SH3 domains clamp the kinase domain by binding the phosphorylated tail and the SH2--kinase linker.\citep{beltran2026src}
Activation releases this clamp, turns $\alpha$C inward and opens the activation loop, a process that MD simulations resolve through metastable intermediates\citep{shukla2014src} into an active basin of about $0.1$--$0.2\%$.\citep{shukla2014src}
Because $\alpha$C and the activation segment rearrange upon activation, we ask whether Boltz-2 resolves both the inactive crystal 2SRC and the active-like crystal 1Y57. To this end, we bias the C$\alpha$ RMSDs of both elements to each crystal after core alignment.

As shown in \Cref{fig:case-src}A, the model landscape resolves a single basin, centered $0.6\,\text{\AA}$ from 2SRC and $7.8\,\text{\AA}$ from 1Y57 (stars, small map). The native structure of Boltz-2 is therefore the inactive crystal.
The active-like basin does not appear even as the model landscape rises to $F_\theta=30$.
The model also keeps the SH2--SH3 clamp closed, with only a $0.081\%$ chance of release (\SICref{fig:evaluations}B). 
For reference, a combined experiment-simulation effort for the related kinase Hck shows that the clamp opens $25\%$ of the time without the tail phosphate.\citep{yang2010hck,huang2020hck}
These signatures of restricted conformational flexibility therefore show that, as for chignolin, the model over-stabilizes the dominant state.

Despite the lack of conformational flexibility, Boltz-2 fluctuations concentrate in the activation loop and the $\alpha$C helix (\Cref{fig:case-src}C).
These residues show the largest RMSD between the active and inactive crystal structures, and their fluctuations point toward the active state (\SICref{app:experiment}).
Thus, while the model only reproduces the inactive structure and overestimates its stability, the learned conformational flexibility still encodes the direction of activation.

\subsection{Two-Well Systems}
\label{sec:double-well}

\begin{figure*}[!t]
\captionsetup{hypcap=false,skip=4pt,font=footnotesize}
\centering
\includegraphics[width=0.95\textwidth]{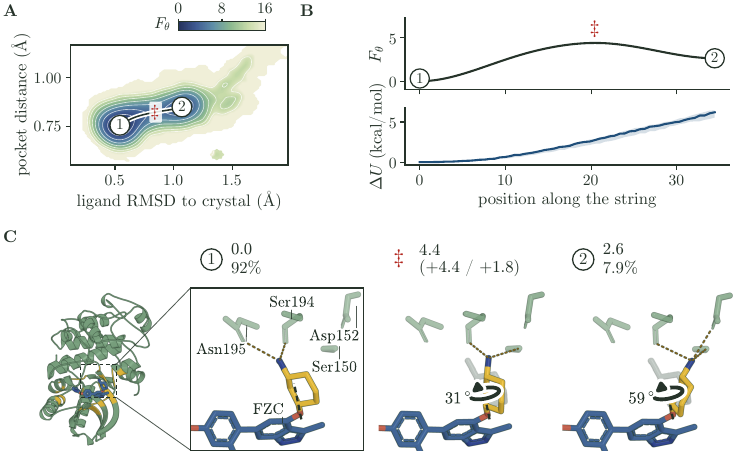}
\caption{MAP2K1--FZC, a double well. (A)~The landscape $F_\theta$ on the ligand RMSD to 7XLP and the pocket distance. (B)~The string profile above the energy $\Delta U$ from adiabatic mapping in ff14SB/GAFF (\SICref{app:string-drag}). (C)~The whole kinase with FZC (slate) and the residues lining it (gold), where the dashed box opens into state~\stmark{1}. The terminal ring is gold over its state-1 position in gray, with its turn about the ether bond (dashed axis) in degrees and dashed lines for the hydrogen bonds of the amine.}
\label{fig:case-double-well}\label{fig:case-mek1}
\end{figure*}

\begin{figure*}[!t]
\captionsetup{hypcap=false,skip=4pt,font=footnotesize}
\centering
\includegraphics[width=0.95\textwidth]{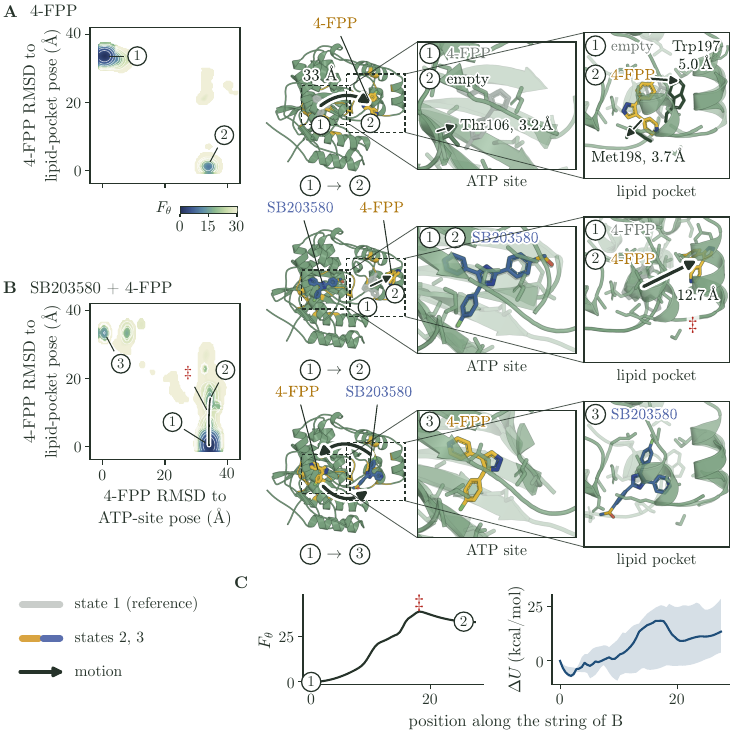}
\caption{p38$\alpha$ in two ligand contexts, on the RMSDs of 4-FPP to its ATP-site and lipid-pocket poses in 2P5A, drawn as in \Cref{fig:case-chignolin}. 
(A)~4-FPP alone and (B)~SB203580 with 4-FPP, each with its landscape, the whole kinase with both pockets (dashed boxes) and close-ups of the two pockets. A rare state is shown in color over state~\stmark{1} in gray, with an arrow for each part that moves (key, bottom left). In (A), 4-FPP moves $33\,\text{\AA}$ from the ATP site into the lipid pocket. In (B), 4-FPP slides $12.7\,\text{\AA}$ out of the lipid pocket over the saddle ($\ddagger$) into state~\stmark{2} (upper row), and state~\stmark{3} exchanges the two ligands on no sampled string, 4-FPP by $33\,\text{\AA}$ and SB203580 by $31\,\text{\AA}$ (lower row). 
(C)~$F_\theta$ along the string of B beside the ff14SB/GAFF energy $\Delta U$ from adiabatic mapping (\SICref{app:string-drag}; curve and dashed region is the median and error bars from multiple initializations of the calculations).}
\label{fig:case-p38}\label{fig:p38-two-pocket}
\end{figure*}

We next turn to two kinase targets with two ligand wells, comparing the sampled poses with experimental structures and evaluating the energetics via adiabatic mapping.

\ResultParagraph{MAP2K1--FZC}
\label{sec:mek1}
We first consider MAP2K1, or MEK1, a cancer-therapy target in the RAF--MEK--ERK chain, which relays growth signals by phosphorylation.
We study its complex with FZC, an ATP-competitive inhibitor that keeps its affinity when MEK is phosphorylated.\citep{takano2023mek} To sample the poses of this ligand, we bias the ligand RMSD to crystal 7XLP and the pocket distance (\Cref{fig:case-mek1}A).
As shown in \Cref{fig:case-mek1}C, the model retains two poses that keep the hinge contacts and differ only in the 3-aminocyclohexyl ring.
The dominant pose ($92\%$ chance) matches the crystal pose to $0.5\,\text{\AA}$, though it is likely memorized since 7XLP precedes the training cutoff (\SICref{tab:experiment}).
In the second pose ($7.9\%$ chance), the ring rotates about its ether linkage, carrying the amine toward Ser150 and Asp152.

Connecting the two states is a string that passes over a saddle (\Cref{fig:case-mek1}B, top).
To validate this double-well picture from the energetics, we perform adiabatic mapping along the string using ff14SB/GAFF (\Cref{fig:case-mek1}B, bottom).
We find that the energy meets no barrier and climbs to $6$~kcal/mol at state~\stmark{2}. The force field thus shows neither the saddle nor the second minimum.
Moreover, the minimized structures display little rotation in the terminal 3-aminocyclohexyl ring.
Combined with the lack of experimental validation on this pose, we may conclude that the second pose is hallucinated.

Although tracing hallucination in a co-folding model is never certain, we note that training data for the MAP2K1 protein also exists in the form of nucleotide-bound structures. 
In fact, two thirds of the nucleotide-bound MEK1 structures place a ribose hydroxyl, a polar group like the amine, at this amine position (\SICref{fig:mek1-amine-sites}).
This result suggests that the hallucinated second pose may emerge from training data over-weighting nucleotide-bound poses, which then dictate the poses of other ATP-site ligands. 

\begin{figure*}[!t]
\captionsetup{hypcap=false,skip=4pt,font=footnotesize}
\centering
\includegraphics[width=\textwidth]{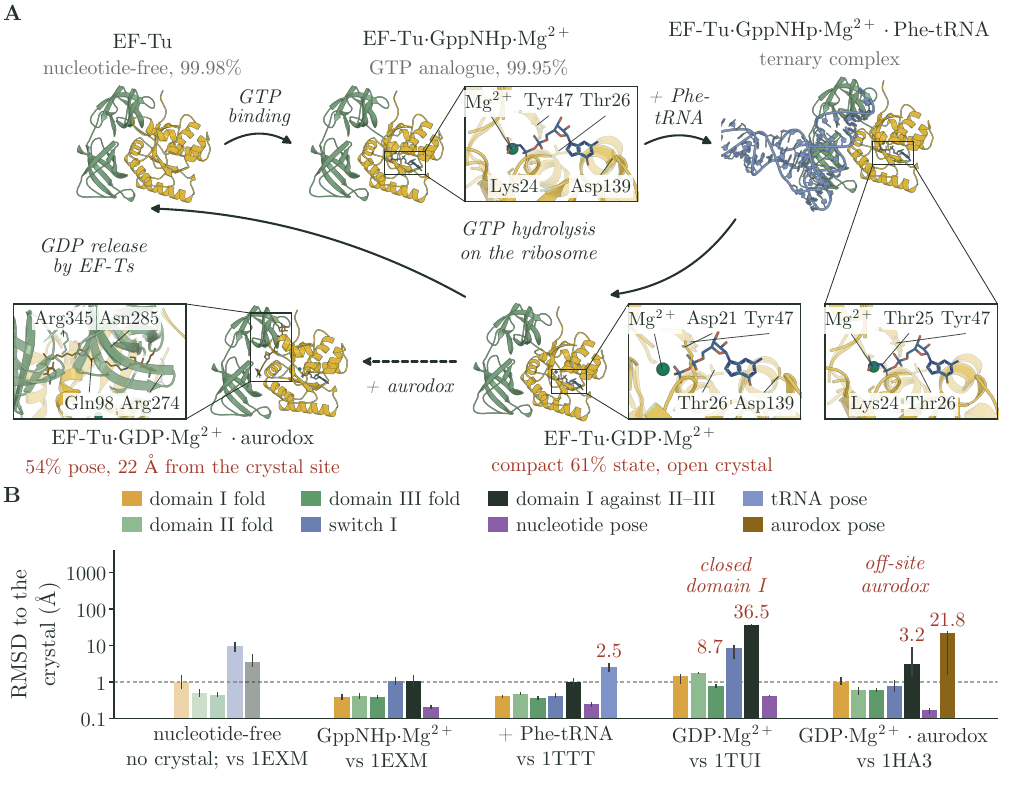}
\caption{The translation cycle with EF-Tu protein. 
(A)~The elongation cycle with each state shown by its most populated generated structure from Boltz-2 after superposing domains II--III (domain~I gold, domains II--III sage, tRNA blue). The framed close-ups enlarge the nucleotide or the antibiotic with its four nearest side chains (\textit{T.~thermophilus} numbering) and Mg$^{2+}$ ion in green. 
Red text marks the two structures that differ significantly from the crystal: the compact GDP form and the aurodox pose. (B)~Accuracy of the independent ensemble of each input against the crystal of its state (median, whiskers $10$--$90\%$; the nucleotide-free protein is set against 1EXM, lighter). Bars give the C$\alpha$ RMSD of each domain superposed on itself, of switch~I after superposing domain~I and of domain~I after superposing domains II--III, and the RMSD of the nucleotide, tRNA and aurodox poses, with the dashed line at $1\,\text{\AA}$ and medians of $2\,\text{\AA}$ or more in red.}
\label{fig:case-eftu}
\end{figure*}

\ResultParagraph{\texorpdfstring{p38$\alpha$}{p38 alpha}}
\label{sec:p38}
We next consider p38$\alpha$, a kinase that mediates inflammatory signaling and contains two drug-binding sites, the ATP cleft and a C-terminal lipid pocket.\citep{perry2009p38}
Two ligands probe these sites. The first, SB203580, is a pyridinylimidazole anti-inflammatory compound selective for p38$\alpha$ over related MAP kinases, and it occupies the ATP cleft.\citep{wang1998p38}
The second, the pyrazolylpyridine lead 4-FPP, binds the same cleft while opening the C-terminal cap into the lipid pocket.\citep{perry2009p38}
In crystal 2P5A (3HVC), 4-FPP occupies both sites, whereas SB203580 fills the ATP site in 1A9U.\citep{wang1998p38,perry2009p38}
To test how ligand context shapes the ensemble of pockets sampled from Boltz-2, we run two variants of p38$\alpha$: (1) p38$\alpha$ with 4-FPP alone, and (2) p38$\alpha$ with SB203580 plus 4-FPP. Both runs bias the two RMSDs of 4-FPP to its ATP-site and lipid-pocket poses in 2P5A.

\begin{figure*}[!t]
\captionsetup{hypcap=false,skip=4pt,font=footnotesize}
\centering
\includegraphics[width=\textwidth]{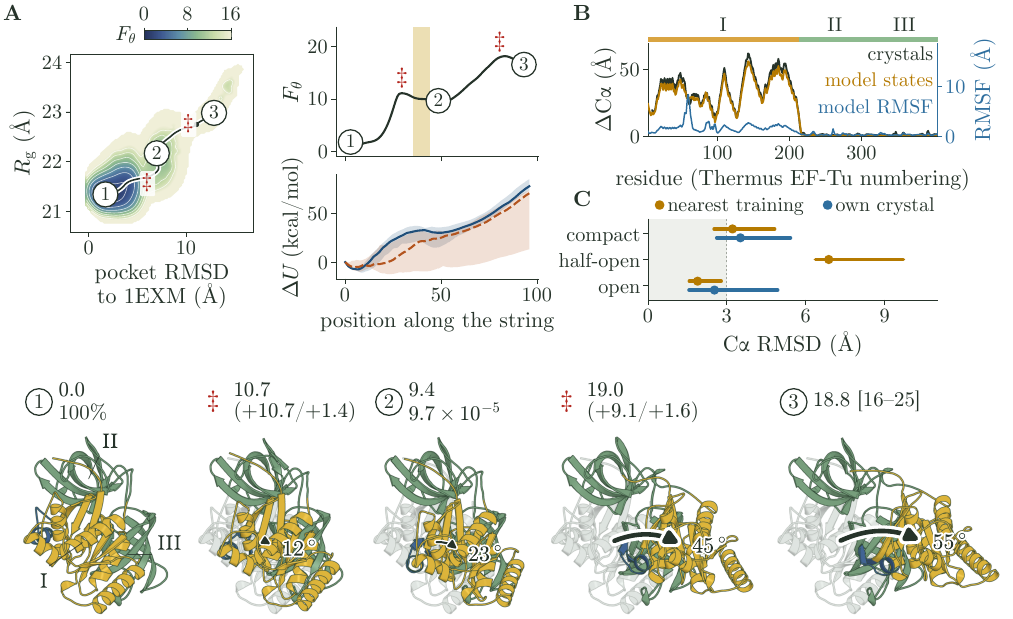}
\caption{Enhanced TPS applied to the nucleotide-free protein. (A)~Landscape of the nucleotide-free protein on the pocket RMSD to 1EXM (protein atoms of the GppNHp crystal) and the radius of gyration, with the profile along the path above the energy $\Delta U$ from adiabatic mapping (\SICref{app:string-drag}; ff14SB solid, median and quartiles of eight starts; AMARO dashed, median and range of four starts). The lower row visualizes the structures in path order (domain~I gold and switch~I slate over domain~I of the first structure in gray; the curved arrow gives its swing); $\ddagger$ marks the two transition states. 
(B)~Per-residue C$\alpha$ displacement (\textit{T.~thermophilus} numbering) between the crystals (black) and between the model's compact and open states (ochre) after aligning domains II--III, with the model's RMSF in blue (right axis). 
(C)~Domain-I C$\alpha$ RMSD of each state's independent structures to the nearest pre-cutoff training chain (ochre) and to the crystal of that state (blue; the half-open state has none), as medians with $10$--$90\%$ ranges over the $3$~\AA\ band.}
\label{fig:case-eftu-apo}
\end{figure*}

As shown in \Cref{fig:case-p38}A, Boltz-2 recovers the 2P5A ATP-site pose at $99.97\%$ chance, with median RMSD $0.61\,\text{\AA}$, against $0.016\%$ in the lipid pocket with median RMSD $1.19\,\text{\AA}$.
Adding SB203580 reverses this preference (\Cref{fig:case-p38}B), filling the ATP site at $99.96\%$ with median RMSD $1.4\,\text{\AA}$ while shifting 4-FPP to the lipid pocket at $99.99\%$ chance, with median RMSD $0.52\,\text{\AA}$.
In contrast, placing 4-FPP at the ATP site is rare with $0.0024\%$ chance (state~\stmark{3}; \Cref{fig:case-p38}B).
In state~\stmark{3} of the two-ligand system, 4-FPP exchanges pockets with SB203580 (\Cref{fig:case-p38}B).
These results demonstrate how Boltz-2 can take diverse ligand context to predict docking at the right pocket. 

Despite correct docking, the lipid pocket itself responds imperfectly to 4-FPP. 
Experimentally, this pocket accepts 4-FPP with or without the Trp197 flip.\citep{perry2009p38}
In the model, Trp197 stays in its closed rotamer with 4-FPP alone and has a $73\%$ chance of staying closed with SB203580, even with 4-FPP in the lipid pocket.
The flipped rotamer, with the remaining $27\%$ chance, brings a several-fold rise in the fluctuations of Trp197 and Met198 (\SICref{fig:p38-rmsf}).
However, the model misses the $\alpha\text{1L}/\alpha\text{2L}$ helix shift at residues $240$--$260$, which moves $1.5$--$2.5\,\text{\AA}$ across crystals and only about $0.2\,\text{\AA}$ in the model.
Boltz-2 thus fits 4-FPP into the lipid pocket without the cap motion of the crystals and gives only a partial response to the ligand.

Lastly, we turn to the energetics of these states.
For the 4-FPP and SB203580 case, state \stmark{1} can be joined to a metastable state \stmark{2} with one string, along which 4-FPP slides out of the lipid pocket (\Cref{fig:case-p38}B).
To validate this state, we perform adiabatic mapping along the string in ff14SB/GAFF (\SICref{app:string-drag}). 
As shown in \Cref{fig:case-p38}C, the force field reproduces the shape of the model profile, where the complex leaves state \stmark{1} over a real energy barrier at $18$~kcal/mol and lands in a shallow basin.
With the saddle as the divide between the two states, the adiabatic-mapping energies alone give state~\stmark{2} a chance of $5\times10^{-10}\%$, against $8\times10^{-13}\%$ from the same string on the model landscape, i.e., well within the tolerance for qualitative comparisons. 
Thus, the force-field energetics corroborate the pathway learned by Boltz-2. 

Overall, p38$\alpha$ is the richest of the two-well systems for Boltz-2, which places each ligand in its crystal pocket with validated energetics. 
These poses nonetheless belong to the training set (1A9U and 3HVC), and the SB239063 plus 4-FPP pose matches 4FA2, another training structure (\SICref{tab:experiment}).
The only out-of-distribution prediction from Boltz-2 is the swap of the two ligands (state~\stmark{3}), which the model predicts with vanishingly small probability. %

\subsection{The Case of EF-Tu}
\label{sec:eftu}

Having seen landscapes with a few conformational basins, we now ask how well a co-folding model describes a complete biological cycle.
We focus on EF-Tu, the GTPase that brings each aminoacyl-tRNA to the ribosome during the elongation step of bacterial protein synthesis.
In its GTP-bound form, EF-Tu delivers the aminoacyl-tRNA to the ribosome, where codon recognition triggers GTP hydrolysis.
In its GDP-bound form, EF-Tu releases the tRNA and leaves the ribosome, which requires the domains to open and the switch~I loop to refold from a helix into a $\beta$-hairpin.
The exchange factor EF-Ts then replaces GDP with GTP, which resets EF-Tu to its starting form.\citep{kjeldgaard1993eftu,berchtold1993eftu,hilgenfeld2000eftu,polekhina1996eftu}
EF-Tu is also the target of the elfamycin antibiotics, including aurodox, which impair its function and have drawn renewed interest because of emerging resistance to approved antibiotics.\citep{prezioso2017elfamycins}

We follow this cycle through five inputs (\Cref{fig:case-eftu}A), namely the nucleotide-free protein, the GTP analogue GppNHp (GNP), GNP with Phe-tRNA (the ternary complex), GDP, and GDP with the antibiotic aurodox. 
By comparing the median RMSD with its crystal (\Cref{fig:case-eftu}B), we find that Boltz-2 reproduces only part of this cycle. 
The median RMSD for GNP and ternary structures lie within about $1\,\text{\AA}$ of their crystals except in the tRNA body. 
With GDP, however, domain~I lies $36.5\,\text{\AA}$ from its crystal position, and with aurodox the antibiotic misses its site by $21.8\,\text{\AA}$. 
Thus, the model predicts the GTP-bound form and the ternary complex well, but misses structural domains of the GDP form and the aurodox binding site.

\begin{figure*}[p]
\captionsetup{hypcap=false,skip=4pt,font=footnotesize}
\centering
\includegraphics[width=0.9\textwidth]{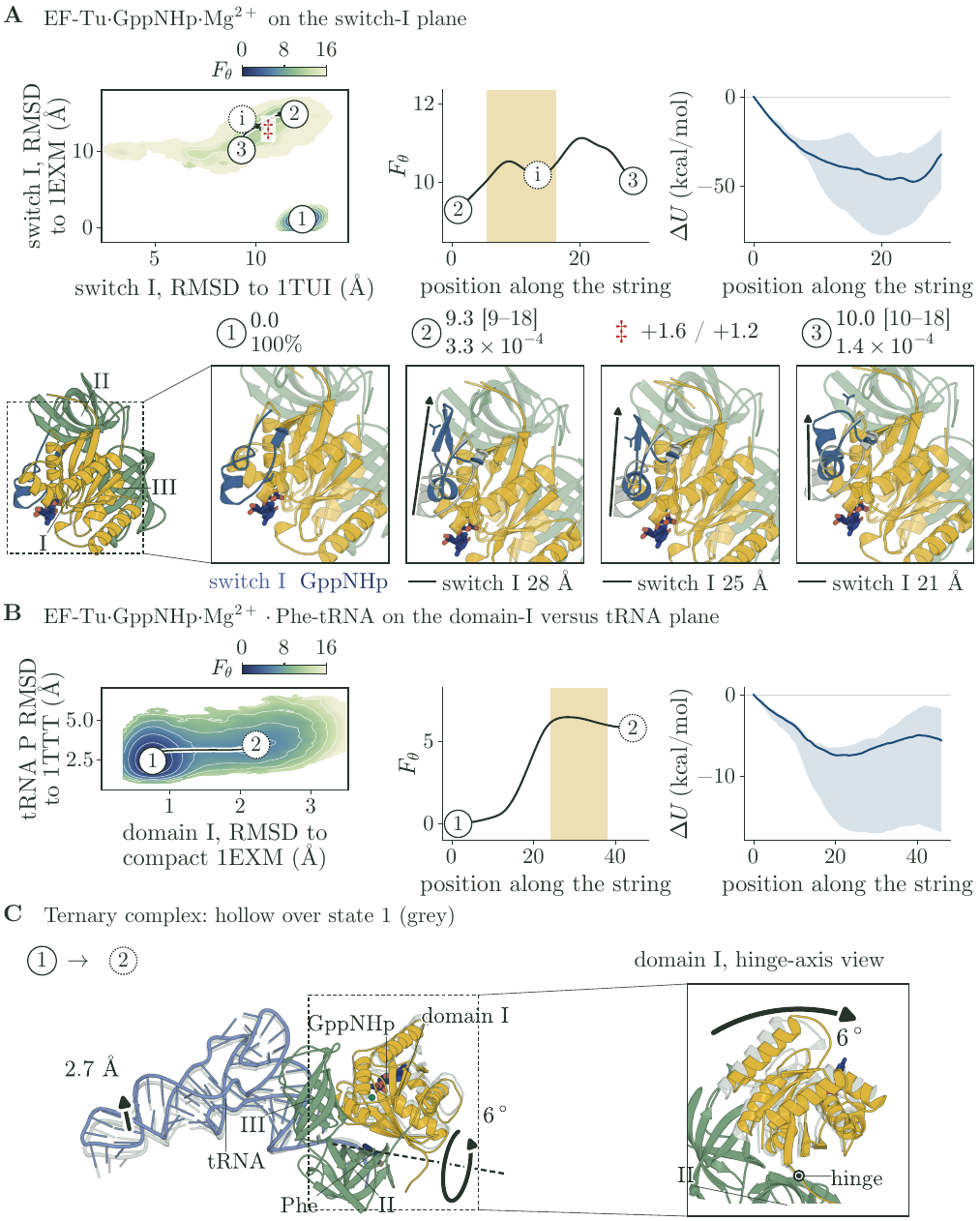}
\caption{Enhanced TPS on the GTP branch, where GppNHp fixes the switch, and Phe-tRNA fixes the domains. (A, B)~(left) The model landscape $F_\theta$ of each input with states, saddles ($\ddagger$) and string, along with the string profile (middle) and the ff14SB/GAFF energy $\Delta U$ profile (right) from adiabatic mapping (median and spread over starts; \SICref{app:string-drag}). 
(A)~GppNHp/Mg$^{2+}$ on the switch-I plane, with the closed switch of 1EXM \stmark{1} and two rare displaced-switch states (\stmark{2} and \stmark{3}). Below, state~\stmark{1} with domain~I in gold, switch~I in slate, and GppNHp in dark blue; the dashed box is enlarged in the first frame. 
The other frames show this site over the gray switch of state~\stmark{1} after superposing domain~I, with an arrow for the residue that moves farthest. 
(B)~The ternary complex on the domain-I versus tRNA plane, with one state \stmark{1} and a shallow second state (\stmark{2}, dotted). (C)~State~\stmark{2} over state~\stmark{1} in gray after superposing domains II--III, with the hinge axis of domain~I (dashed, $6^\circ$ turn), the shift of the anticodon arm (arrow, enlarged fourfold) and a close-up along this axis.}
\label{fig:case-eftu-gtp}
\end{figure*}

\begin{figure*}[!t]
\captionsetup{hypcap=false,skip=4pt,font=footnotesize}
\centering
\includegraphics[width=\textwidth]{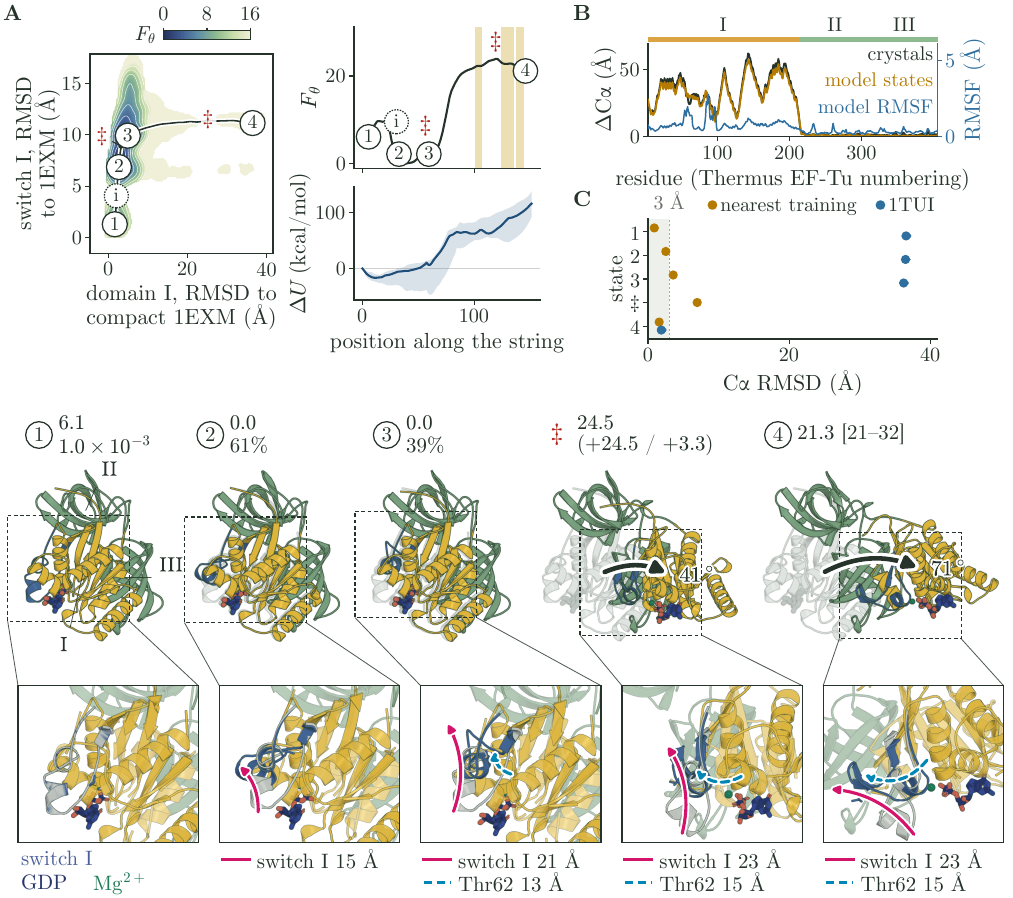}
\caption{Enhanced TPS on EF-Tu with GDP, where the protein stays compact. 
(A)~Landscape of GDP/Mg$^{2+}$ on the domain-I versus switch-I plane, with the string profile above the ff14SB/GAFF energy $\Delta U$ from adiabatic mapping (\SICref{app:string-drag}; median and quartiles of eight starts). Below, the structures in path order in the view of \Cref{fig:case-eftu-apo} (domain~I gold and switch~I slate over domain~I of the first structure in gray; GDP dark blue; curved arrow, swing of domain~I). Each dashed box is enlarged in the frame below, over the switch of state~\stmark{1} in gray after superposing domain~I (Mg$^{2+}$ green), with arcs for the switch-I residue that travels farthest (magenta, solid) and for the Thr62 hydroxyl (cyan, dashed). 
(B)~Per-residue C$\alpha$ displacement between the crystals 1EXM and 1TUI (black) and between states 2 and 4 (ochre), with the RMSF of state~\stmark{2} (blue, right axis). 
(C)~Domain-I C$\alpha$ RMSD of each state to the nearest pre-cutoff training chain (ochre) and to 1TUI (blue, states only), as medians with $10$--$90\%$ ranges.}
\label{fig:case-eftu-gdp}
\end{figure*}

\begin{figure*}[p]
\captionsetup{hypcap=false,skip=4pt,font=footnotesize}
\centering
\includegraphics[width=\textwidth]{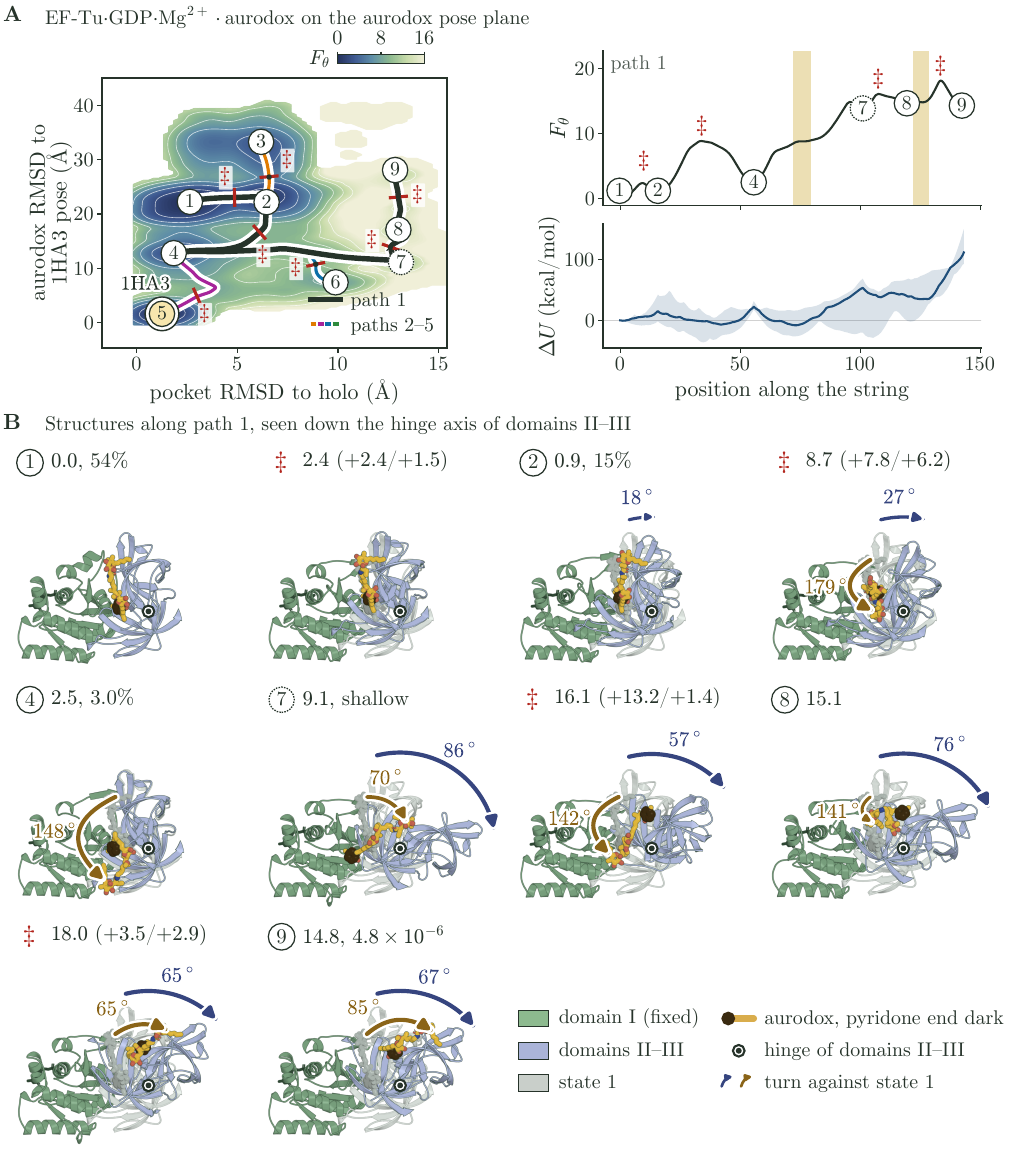}
\caption{Enhanced TPS on EF-Tu with GDP and aurodox along path~1 of the aurodox pose plane. 
(A)~The model landscape $F_\theta$ with path~1 (heavy line), paths~2--5 (colors) and their transition states (red bar across the path, $\ddagger$ beside it), and $F_\theta$ along path~1 above the ff14SB/GAFF energy $\Delta U$ from adiabatic mapping (\SICref{app:string-drag}; median and range of four starts). State~\stmark{5}, the pose of the crystal 1HA3, carries a pale gold circle inside a second ring. 
(B)~The structures in string order, superposed on domain~I and seen down the hinge axis of domains~II--III, over state~\stmark{1} in gray. Arrows give the turn against state~\stmark{1} of domains~II--III (slate) and of the antibiotic (gold). Each transition state carries its $F_\theta$ and, in brackets, its heights above the two neighboring states. 
The antibiotic swings by $148^\circ$ about its pyridone end from domain~II into the cleft of domains~I and~III (4). Domains~II--III then turn by $57^\circ$--$86^\circ$ about one hinge (7--9), and in state~\stmark{9} they carry the antibiotic at its state-1 site. The aurodox structures are non-validated configurational isomers (\SICref{si:diag-amber}).}
\label{fig:case-eftu-aurodox}
\end{figure*}

\begin{figure*}[p]
\captionsetup{hypcap=false,skip=4pt,font=footnotesize}
\centering
\includegraphics[width=\textwidth]{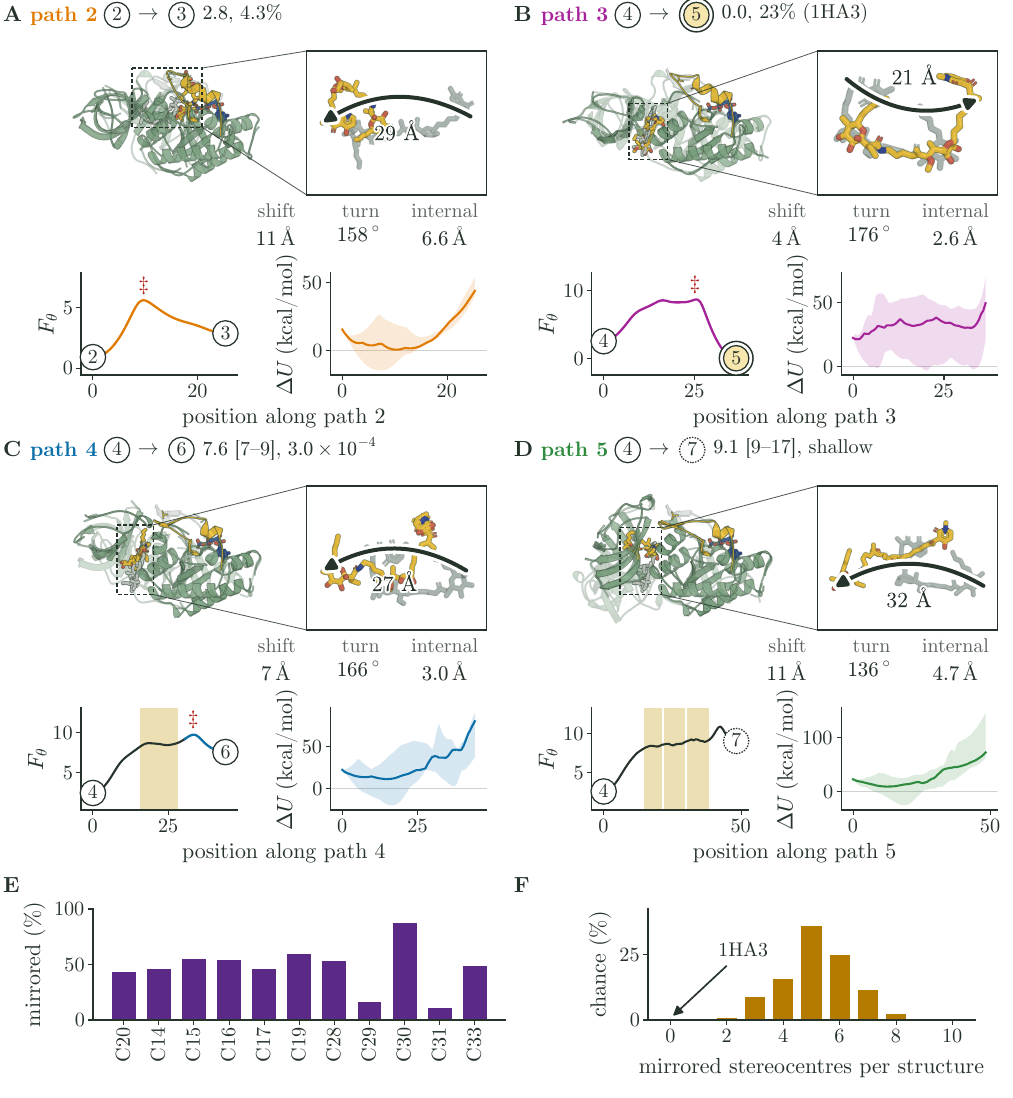}
\caption{EF-Tu with GDP and aurodox, where the antibiotic turns end over end along every branch. 
(A--D)~Paths 2--5 of the aurodox pose plane (\Cref{fig:case-eftu-aurodox}A), each with its last state over the pose of its first state (gray), the antibiotic alone in both states (arrow, the atom that travels farthest; below, centroid shift, turn and internal RMSD), and the profile of the path beside the ff14SB/GAFF energy $\Delta U$ from adiabatic mapping (\SICref{app:string-drag}; median and range of four starts). Paths 2--4 each cross one transition state ($\ddagger$), whereas path~5 ends in a shallow hollow (7) without a transition state. A stretch shared with path~1 is black, and $\Delta U$ continues from path~1. Path~3 ends in state~\stmark{5}, the pose of the crystal 1HA3 (pale gold circle inside a second ring). 
(E, F)~Handedness of the generated antibiotic relative to the 1HA3 ligand: the chance that each stereocenter comes out mirrored (E) and the chance of each number of mirrored stereocenters in one structure (F), with the crystal ligand at zero (\SICref{si:diag-amber}).}
\label{fig:case-eftu-aurodox-branches}
\end{figure*}

Having established the general accuracy, in what follows, we now examine each system and ask whether the ensemble follows the biochemistry of its step in the cycle (\Cref{fig:case-eftu-apo,fig:case-eftu-gtp,fig:case-eftu-gdp,fig:case-eftu-aurodox,fig:case-eftu-aurodox-branches}; \SICref{methods:eftu-context}).

\paragraph{Nucleotide-free protein.}
Our analysis begins with the nucleotide-free EF-Tu, which is the intermediate structure during the nucleotide exchange when EF-Ts binds EF-Tu and disrupts its Mg$^{2+}$ site to release GDP.\citep{kawashima1996efts}
As the only input without a ligand and without a crystal structure of its own, this study probes the ensemble that the model assigns to the protein alone, without nucleotide or Mg$^{2+}$ ion.
To sample this ensemble, we bias the RMSD of the aurodox-site pocket to 1EXM and the radius of gyration (\Cref{fig:case-eftu-apo}).
To validate the resulting structures, we compare them with the nucleotide-bound crystals, namely the open GDP form 1TUI and the compact GppNHp form 1EXM, calculating the RMSD only over protein atoms.

As shown in \Cref{fig:case-eftu-apo}A (left), the model generates three conformational basins joined by a single string between the compact and open forms.
The compact region is observed with $99.98\%$ probability, while the half-open intermediate and open forms appear with $0.018\%$ and $0.003\%$ probability.
The open form agrees with crystallography, with a median C$\alpha$ RMSD of $1.2\,\text{\AA}$ from the GDP form 1TUI (\Cref{fig:case-eftu-apo}B).
However, nearly all open structures lie within $3\,\text{\AA}$ of 1TUI or of the \textit{M.~tuberculosis} GDP complex 7VOK, both in the training set, which makes memory recall a likely explanation for this high accuracy.

The half-open intermediate has no counterpart in the training set. In this state (\Cref{fig:case-eftu-apo}A, lower row), switch I has formed its GDP-like hairpin while the G domain has turned only partway; no such structure is found in the training set (\Cref{fig:case-eftu-apo}C).
To validate this intermediate, we perform adiabatic mapping along the path in ff14SB and in AMARO (\Cref{fig:case-eftu-apo}A, lower profile).
The ff14SB energy peaks shortly before the intermediate and dips into a minimum exactly at this state.
In contrast, the AMARO energy rises through this state before an inflection change at the state \stmark{2} location. 
Beyond this state, both energies rise steadily to the open form but since the open form matches the crystal 1TUI, its energy minimum may lie beyond the end of the string.

In summary, Boltz-2 can identify the correct closed and open states. 
The intermediate state has no match from within the training set, but can be viewed as a valid state from the adiabatic mapping calculations. 
This result suggests that Boltz-2 has learned more than a recall of the two crystal forms, since it connects these training-set states through an unseen intermediate supported by force-field energetics.

\paragraph{GTP analogue.}
With the protein alone characterized, we now turn to its complex with the GTP analogue GppNHp, the next step of the cycle.
GTP binding activates EF-Tu. In the GppNHp crystal 1EXM, the domains are compact and switch I is a helix whose Thr62 binds the Mg$^{2+}$, which is the form that binds aminoacyl-tRNA.\citep{kjeldgaard1993eftu,berchtold1993eftu} To sample this complex, we bias the two C$\alpha$ RMSDs of domain~I after superposing domains II--III, one to the compact form 1EXM and one to the open form 1TUI (\SICref{tab:cvs}). Since 1EXM is now the crystal of the input itself, we then analyze the ensemble on the switch-I plane of \Cref{fig:case-eftu-gtp}A, which is measured after superposing domain~I and carries no bias.

As shown in \Cref{fig:case-eftu-gtp}A, the model responds to GNP, with switch I having a $99.95\%$ chance of lying within $3\,\text{\AA}$ of its 1EXM geometry, against a $0.15\%$ chance by the same measure in the previous nucleotide-free case.
The remaining $0.05\%$ occupy two displaced-switch states (2, 3) in which Thr62 detaches from Mg$^{2+}$, joined by a string over a low saddle.
Experimentally,single-molecule FRET finds free EF-Tu in solution to have more conformations than the two crystal forms, spanning a continuum of domain arrangements between the compact and open crystals.\citep{johansen2018eftu}
With GNP, by contrast, the model holds domain~I within about $1.5\,\text{\AA}$ of its 1EXM position at $90\%$ chance (\Cref{fig:case-eftu}B).
The model ensemble with GNP is thus narrower than the spread in solution.
With that being said, we note that adiabatic mapping from state~\stmark{2} to state~\stmark{3} shows that the ff14SB/GAFF energy falls by a median of $32$~kcal/mol without a barrier, whereas the model profile stays nearly flat (\Cref{fig:case-eftu-gtp}A, right).
Thus, from the standpoint of force-field energetics, there is no support for these rare states, which implies that Boltz-2 did not sample the right metastable conformations of this complex.

\paragraph{Ternary complex.}
The crucial state in the translation cycle is the ternary complex of EF-Tu, GTP, and aminoacyl-tRNA.
This complex carries the amino acid to the ribosome, and the tRNA contacts all three domains of EF-Tu~\citep{nissen1995ternary,voorhees2010activation}.
To sample this complex, we again bias the two domain-I RMSDs to 1EXM and 1TUI (\SICref{tab:cvs}).
We then project the reweighted ensembles onto an unbiased tRNA coordinate, the RMSD of the tRNA phosphorus atoms from the crystal 1TTT (\Cref{fig:case-eftu-gtp}B).

Binding of Phe-tRNA restricts the domain arrangement of EF-Tu to a single basin on every plane (\Cref{fig:case-eftu-gtp}B), narrowing the ensemble further relative to the nucleotide-free and GNP cases.
The tRNA contacts all three domains in this basin as in the crystal structure 1TTT~\citep{nissen1995ternary}, with Phe and the CCA end in the domain-II pocket at the interface with domain~I, the T stem packed on domain~III, and the acceptor stem touching domain~I.
The representative structure of state~\stmark{1} places Phe against the same ten residues as the crystal, His67 of domain~I and nine residues of domain~II.
Enhanced sampling finds only one additional, rare conformational basin (state~\stmark{2}), reached by small rigid-body turns of domain~I about its hinge with domain~II and of the tRNA body about its acceptor stem (\Cref{fig:case-eftu-gtp}C).
This state lies closer to the ribosome-bound complex 4V5L~\citep{voorhees2010activation} than to the free crystal ($1.8$ against $3.0\,\text{\AA}$), suggesting that it reflects the presence of 4V5L in the training data.

Moving on to the energetic evaluation, we find that adiabatic mapping with ff14SB/GAFF invalidates the double-well picture (\Cref{fig:case-eftu-gtp}B, right).
Instead of crossing a saddle point, the energy profile falls to $-6$~kcal/mol halfway along the string and then rises by about $2$~kcal/mol toward state~\stmark{2}.
Furthermore, heavy-atom overlaps below $1.8\,\text{\AA}$ appear in $96\%$ of the generated structures scored for this input.
These overlaps are few and local, since each representative structure holds only one to four compressed polar contacts among about $4{,}800$ heavy atoms (\SICref{si:diag-amber}).
Thus, although Boltz-2 samples the ternary complex with the right structural arrangement, its predicted population ratios are unreliable.

\paragraph{GDP.}
With the GTP-bound branch characterized, we now turn to the GDP branch, where the biochemistry of release from the ribosome requires domain opening.
Indeed, in the GDP crystal 1TUI, the G domain has rotated away and switch I is found in a hairpin state.\citep{polekhina1996eftu}
This open form has also been corroborated with single-molecule FRET in solution.\citep{johansen2018eftu}
To follow both the domains and the switch, we analyze the ensemble on the plane of \Cref{fig:case-eftu-gdp}A, which pairs the biased domain-I RMSD to 1EXM with the switch-I RMSD.

In contrast, Boltz-2 predicts closed domains in the dominant state, as domain I has a $0.0009\%$ chance of lying within $10\,\text{\AA}$ of the open reference (\SICref{fig:evaluations}B).
GDP changes only the switch through a series of four connected states (\Cref{fig:case-eftu-gdp}A).
The string starts at state~\stmark{1}, the closed switch of the GTP form in which Thr62 binds the Mg$^{2+}$, with only a $0.10\%$ chance of being observed.
In state~\stmark{2}, the switch has left this position but keeps its helix, with a $61\%$ chance.
In state~\stmark{3}, with the remaining $39\%$, the switch gives up this helix, and the hairpin forms only in this state.
From state~\stmark{3}, the string crosses a half-open saddle and ends at the open state~\stmark{4} (\Cref{fig:case-eftu-gdp}C).
When we reach state~\stmark{4}, the displacement of each residue relative to state~\stmark{2} matches exactly the displacement between 1EXM and 1TUI (\Cref{fig:case-eftu-gdp}B).
Thus, while the population ratios underestimate the open form, its structure is still predicted at high accuracy.

To test these states, we again perform adiabatic mapping along the string in ff14SB/GAFF (\Cref{fig:case-eftu-gdp}A, lower profile).
The energy stays within about $16$~kcal/mol from state~\stmark{1} through state~\stmark{3} and then climbs steeply to $116$~kcal/mol at state~\stmark{4}, which qualitatively matches the string profile.
However, the minimized complex does not reach the open structure sampled by Boltz-2.
Adiabatic mapping fixes only two RMSDs, and the complex that it produces at state~\stmark{4} differs from the open structures of the model by $6.8\,\text{\AA}$.
The $116$~kcal/mol is therefore the energy of this other structure and says little about the open state itself.
To obtain the energy of the open state, we instead take the structures sampled by Boltz-2 in each state and minimize them directly in the same force field.
With these structures, state~\stmark{4} lies only $8$~kcal/mol above states~\stmark{2} and~\stmark{3}, a difference smaller than the spread within a single state.
Thus, the force field does not penalize the open form, in agreement with experiment, and the rarity of this form belongs to the model alone.

\paragraph{Aurodox.}
Finally, we end our EF-Tu case studies with aurodox.
As a kirromycin-family antibiotic, it holds EF-Tu$\cdot$GDP in its GTP-like form, which is thought to fix the factor on the ribosome.
With the antibiotic bound, EF-Tu$\cdot$GDP stays on the ribosome after hydrolysis and stops protein synthesis.\citep{prezioso2017elfamycins}
In the crystal 1HA3, aurodox sits within the cleft between domains I and III.\citep{vogeley2001aurodox}
To resolve the poses of the antibiotic, we analyze the ensemble on the pose plane of \Cref{fig:case-eftu-aurodox}A, which pairs the pocket RMSD with the RMSD of the antibiotic to 1HA3 and carries no bias.
Of all protein--ligand complexes in this study, the aurodox complex has the richest landscape, with ten states and five strings on the plane of the aurodox pose (\Cref{fig:case-eftu-aurodox}A).
However, the dominant state~\stmark{1}, with a $54\%$ chance, places the antibiotic against domain~II, $22\,\text{\AA}$ from its site in 1HA3.
The crystal pose is instead sampled at state~\stmark{5}, which lies $1.7\,\text{\AA}$ from 1HA3 and has a $23.3\%$ chance of being observed.

The longest string, path~1, runs from state~\stmark{1} to the rare open states~\stmark{8} and~\stmark{9}, where the antibiotic first swings from domain~II into the cleft of domains~I and~III (state~\stmark{4}).
Domains~II--III then turn about a single hinge axis, which opens the protein and carries the antibiotic back to its state-1 site (\Cref{fig:case-eftu-aurodox}B).
Along path~1, four shorter strings branch from path~1 at states~\stmark{2} and~\stmark{4} (\Cref{fig:case-eftu-aurodox-branches}A--D).
Among these branches, path~3 reaches state~\stmark{5}, so the model connects its dominant pose to the crystal pose.

Under adiabatic mapping in ff14SB/GAFF, the complex climbs a median of $49$~kcal/mol from the dominant pose to the crystal pose.
Along path~1, the adiabatic-mapping energy rises with $F_\theta$ from the dominant pose to state~\stmark{9}, and both profiles order the states nearly alike (\Cref{fig:case-eftu-aurodox}A, right).
It also peaks just after three of the four saddles of the model.
On every branch, by contrast, the adiabatic-mapping energy is highest at the last state, and on path~2 it even dips at the saddle of the model (\Cref{fig:case-eftu-aurodox-branches}A--D).
Path~3 descends in $F_\theta$ to the crystal pose, whereas the minimized complex climbs to this pose (\Cref{fig:case-eftu-aurodox-branches}B).
Thus, apart from the qualitative agreement in path~1, the branching pathways learned by the model are not reflected in the energies.

In addition to this conformational flexibility, the generated antibiotic also displays stereochemical flexibility. 
In aurodox, there are eleven stereocenters whose handedness can be read from the sampled pose and compared with that of the crystal ligand in 1HA3 (\SICref{si:diag-amber}).
As shown in \Cref{fig:case-eftu-aurodox-branches}E, eight of the eleven centers have a chance near $50\%$ of coming out mirrored. One center is mirrored with an $87\%$ chance, and only two centers keep the crystal's handedness in most structures.
In other words, a typical aurodox configuration would carry five mirrored centers (\Cref{fig:case-eftu-aurodox-branches}F). 
Thus, each pose sampled from Boltz-2 belongs to one of many stereoisomers of aurodox.

Experiments, by contrast, report no such flexibility.
Aurodox was assigned a single configuration at each of its eleven stereocenters,\citep{maehr1980aurodox} which the crystal ligand of 1HA3 shares at every center.\citep{vogeley2001aurodox}
While related antibiotics may form epimers at the carbon next to the amide carbonyl in mild base, aurodox remained a single compound under the same conditions.\citep{carter1995isomerism}
Boltz-2 mirrors this very carbon center in $53\%$ of its structures, so the stereochemical flexibility is a failure of the model with no counterpart in experiment.
Note that enhanced TPS draws from the learned distribution alone, without the inference-time potentials that Boltz-2 offers for enforcing stereochemistry.\citep{passaro2025boltz2}

Overall, aurodox is the most complex ligand case study. 
Boltz-2 holds the crystal pose as its second state and connects it to the rest of the landscape, yet it prefers a pose on domain~II far from the crystal site.
The force field follows the rise of path~1 but disagrees with the model on every branch.
Lastly, the landscape holds more poses than the single crystal structure, where the model proliferates a range of mirrored stereoisomers.

\begin{figure*}[!tp]
\captionsetup{hypcap=false,skip=4pt,font=footnotesize}
\centering
\includegraphics[width=\textwidth]{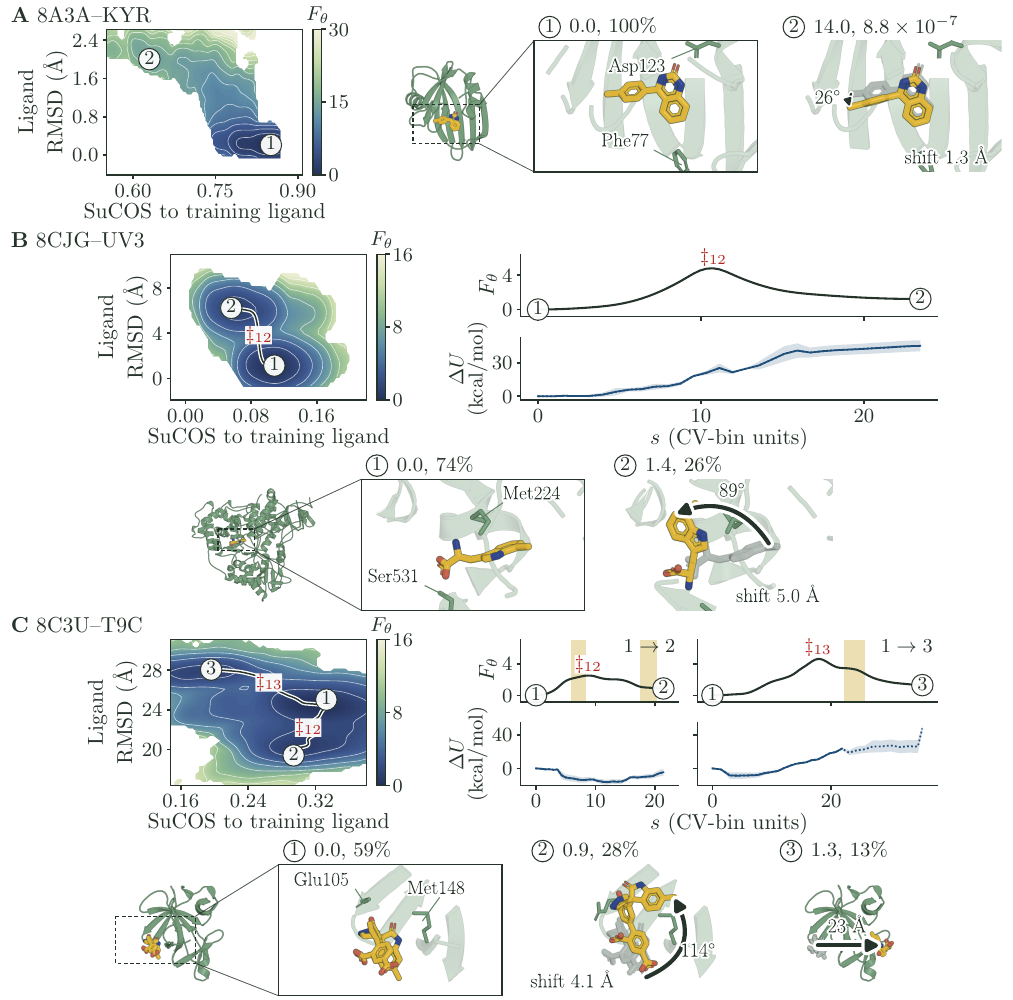}
\caption{Selected protein-ligand complexes from the Runs N' Poses benchmark: 8A3A--KYR (A), 8CJG--UV3 (B) and 8C3U--T9C (C). Each landscape is drawn on the ligand RMSD from the crystal pose against SuCOS, with states (circles) and ridges ($\ddagger$). For 8CJG and 8C3U, it stands beside $F_\theta$ along the string coordinate $s$ (shaded bands are plateaus) and the ff14SB/GAFF energy $\Delta U$ from adiabatic mapping with both coordinates held (median and interquartile range over two starts, dotted where the score stays more than $10^{-2}$ from the string; \SICref{app:string-drag-exact}). The structures show the whole protein with the pocket of state~\stmark{1} (dashed box), its close-up (frame) and each further state, headed by its $F_\theta$ and chance, over the gray pose of state~\stmark{1}, with the turn and the centroid shift of the ligand. The two states of 8A3A have no sampled path between them, and state~\stmark{3} of 8C3U binds at a second site.}
\label{fig:case-rnp}\label{fig:case-rnp-kyr}\label{fig:case-rnp-t9c}\label{fig:case-rnp-uv3}
\end{figure*}

\subsection{Memorization}
\label{sec:experiment}\label{sec:memorization}\label{sec:beyond-benchmarks}

The case studies show that Boltz-2 does not learn the conformational ensemble well. 
Although the model reaches high structural accuracy, it predicts the wrong population ratios in nearly every case, in line with earlier tests of co-folding models.\citep{buttenschoen2024posebusters,masters2026generalization,sun2026kinconfbench,obendorf2026decoupling}
The literature often attributes such errors to memorization of the training data.\citep{skrinjar2026memorization,masters2026generalization,nittinger2025cofolding}
For example, the Runs N' Poses benchmark scores each of its $2{,}600$ post-cutoff complexes by the SuCOS similarity of its crystal ligand to the training set, and finds that the models largely memorize ligand poses.\citep{skrinjar2026memorization}
However, SuCOS was assigned only once per complex and was not used to describe memorization of the poses sampled by a co-folding model.
Inspired by their construction, we use SuCOS as a collective variable for enhanced sampling and, in doing so, test the extent of memorization in the conformational ensemble.

To define this variable, we first recall how SuCOS compares the sampled ligand of a structure $\bm x$ with a training ligand $k$.\citep{leung2019sucos,skrinjar2026memorization}
After overlaying the two ligands, the score combines three terms, the first of which is the feature overlap
\begin{equation}
F_k=\frac{1}{\min(n,n_k)}\sum_{a=1}^{n}\,\sum_{b\in\mathcal B_k(a)}e^{-d_{ab}^2/\sigma^2},
\label{eq:sucos-feature}
\end{equation}
where the outer sum runs over the $n$ pharmacophore features $a$ of the sampled ligand, such as hydrogen-bond donors and acceptors, charged groups, aromatic rings, and hydrophobes, and $n_k$ is the number of features of ligand $k$.
For each feature $a$, $\mathcal B_k(a)$ is the set of features of ligand $k$ that share the type of $a$ and lie within $2.5$~\AA, $d_{ab}$ is the distance of such a pair, and $\sigma=1$~\AA.
Intuitively, a coincident pair of features results in $F_k=1$ and a distant pair contributes nearly zero, while the prefactor normalizes the overlap. The second term is the shape protrusion,
\begin{equation}
P_k=1-\frac{V_{\cap k}}{V},
\label{eq:sucos-shape}
\end{equation}
where $V$ is the volume of the sampled ligand and $V_{\cap k}$ is the volume that it shares with ligand $k$.
Physically, $P_k$ is the fraction of the sampled ligand that lies outside the training ligand.
A pose buried in the training ligand thus gives $P_k=0$, whereas a pose that misses the training ligand entirely gives $P_k=1$.
The third term is the pocket coverage,
\begin{equation}
c_k=\frac{|\mathcal R(\bm x)\cap\mathcal R_k|}{|\mathcal R(\bm x)|},
\label{eq:sucos-pocket}
\end{equation}
where $\mathcal R(\bm x)$ is the set of residues within $6$~\AA\ of the sampled ligand and $\mathcal R_k$ is the set of residues aligned to the pocket of ligand $k$.
This term discounts a training ligand that binds at another site of the protein. With $F_k$ and $P_k$ clipped to $[0,1]$, the three terms combine into
\begin{equation}
\mathrm{SuCOS}(\bm x)=\max_k\;c_k\Bigl[\tfrac12F_k+\tfrac12\bigl(1-P_k\bigr)\Bigr],
\label{eq:sucos}
\end{equation}
where the maximum over $k$ assigns each structure the score of its nearest training ligand.
A pose that matches a training ligand in features, shape and site thus scores near one.
Note that Runs N' Poses evaluates \cref{eq:sucos} once, on the crystal structure, whereas here $\bm x$ is any sampled structure.
As a result, SuCOS shows whether, among correct answers, the model moves closer to the training set.

An example application of using SuCOS as a collective variable in enhanced TPS is shown in \Cref{fig:case-rnp} for three post-cutoff complexes of Runs N' Poses.
We chose these systems, whose crystal ligands lie both near and far from the training set, to evaluate how much an ensemble is memorized from the training data.
For each complex, we sample the ensemble with Composite MetaD and project it onto SuCOS and the ligand RMSD from the crystal pose, where a pose within $2$~\AA\ of the crystal ligand counts as roughly correct.
As shown in \Cref{fig:case-rnp}, the three complexes give three different qualitative features. 
\Cref{fig:case-rnp}A shows the model landscape and sampled poses for 8A3A where the predicted pose retains a SuCOS of $0.85$, reflecting the SuCOS value of the crystal ligand pose in 8A3A. 
The model also produces a rare ligand pose \stmark{2}, which is further away from the training set. 
Correctness and memorization are therefore inseparable for this complex.
In contrast, for 8CJG, the correct pose has a $74\%$ chance of being observed with a very low SuCOS (\Cref{fig:case-rnp}B)--- a low score shared by the second conformational basin \stmark{2}.
This result shows that Boltz-2 reaches the correct pose here without moving toward the training set.
For 8C3U, however, the model samples no correct pose (\Cref{fig:case-rnp}C) and all three sampled states lie $20$--$28$~\AA\ from the crystal and score several times higher than the crystal on SuCOS.
The model thus selects the poses that resemble its training set, affecting its ability predict the correct pose.

The same landscapes also show how the poses of each complex connect.
For 8A3A, the landscape holds a single state at the crystal pose (\Cref{fig:case-rnp}A).
The only other pose, state~\stmark{2}, turns the ligand by $26^\circ$ toward Phe77 and has a negligible chance with no sampled path to state~\stmark{1}.
For 8CJG, the string leaves the correct pose of state~\stmark{1} over the saddle $\ddagger_{12}$ and ends in state~\stmark{2} (\Cref{fig:case-rnp}B).
In this state, the ligand has turned by $89^\circ$ and shifted by $5$~\AA.
For 8C3U, two strings leave state~\stmark{1} (\Cref{fig:case-rnp}C).
The first crosses $\ddagger_{12}$ to state~\stmark{2}, where the ligand has turned by $114^\circ$ near the same site.
The second crosses the higher saddle $\ddagger_{13}$ to state~\stmark{3}, at a second site $23$~\AA\ from the first.
From adiabatic mapping of 8CJG and 8C3U in ff14SB/GAFF (\Cref{fig:case-rnp}B, C), we find mixed results. 
For 8CJG, the energy meets no barrier and climbs to $45$~kcal/mol at state~\stmark{2}, so the force field shows neither the saddle nor the second minimum.
For 8C3U, the energy falls below its starting value on the way to state~\stmark{2}, whereas it climbs to $48$~kcal/mol at state~\stmark{3}.

Lastly, we note that these strings do not follow one stereoisomer throughout.
The ligands of 8CJG and 8C3U each have a stereocenter with the reference configuration $S$, and Boltz-2 generates both configurations in every state (\SICref{si:diag-amber}).
The mirrored configuration $R$ has a chance of $60\%$ in state~\stmark{1} of 8CJG and $77\%$ in state~\stmark{1} of 8C3U, against $17$--$41\%$ in the other states.
The structures drawn on the two sides of $\ddagger_{12}$ indeed differ in stereoisomer (\Cref{fig:case-rnp}B, C).
As with aurodox (\Cref{fig:case-eftu-aurodox-branches}E), the model thus samples the handedness of a ligand along with its pose.

Overall, these three complexes show that memorization can be measured, with the right enhanced sampling technique, for a whole ensemble. 
We therefore expect SuCOS, as a collective variable, to extend benchmarks such as Runs N' Poses as an evaluation tool for conformational ensembles. 
Such benchmarks can then guide future co-folding models toward correct populations as well as correct structures.

\section{Discussion}
\label{sec:discussion}\label{sec:conclusion}

In summary, TPS brings enhanced-sampling tools from molecular simulations to co-folding models, allowing us to probe their conformational ensembles. 
Because the method acts only on the sampler, it requires neither retraining the model nor estimating the likelihood. 
Using well-chosen collective variables (CVs), we may use enhanced TPS to measure the underlying model landscape $F_\theta(s)$, which maps the various conformational basins and saddle points learned by the model. 
The case studies show the complexity and unevenness of the conformations learned by the model. 

Applied to Boltz-2, we find that the landscape carries rich structural information that agrees with experiments and with the energies from adiabatic mapping.
For p38$\alpha$, Boltz-2 places each ligand in the right pocket (\Cref{fig:case-p38}).
Adiabatic mapping calculations also reproduce the barrier and the shallow basin along the path of 4-FPP out of the lipid pocket.
For nucleotide-free EF-Tu, enhanced TPS uncovers a transition pathway that opens the protein to within $1.2$~\AA\ of the GDP crystal form through a half-open intermediate (\Cref{fig:case-eftu-apo}).
This intermediate is absent from the training set, yet it is validated as an appropriate energy minimum from the adiabatic mapping calculation. 
Moreover, the fluctuations of Src kinase in its closed form point toward its active form, and the open form of EF-Tu with GDP reproduces the displacement between the two crystal forms residue by residue (\Cref{fig:case-src,fig:case-eftu-gdp}).

However, other results disagree with experiments or with the force field.
In general, Boltz-2 over-stabilizes its dominant states, so that the open form of EF-Tu with GDP is almost unpopulated, although this open form is the crystal structure and carries no penalty from the force-field calculation (\Cref{fig:case-eftu-gdp}).
Furthermore, some states learned by the model lack experimental evidence or support from adiabatic mapping.
For example, adiabatic mapping shows neither a saddle nor a minimum for the second ligand pose in MAP2K1-FZC (\Cref{fig:case-mek1}).
For aurodox, Boltz-2 prefers a pose far from the crystal site and mirrors a median of five of the eleven stereocenters, a variation without counterpart in experiments (\Cref{fig:case-eftu-aurodox-branches}).
Lastly, the ligand poses of 8C3U-T9C resemble the training ligands far more than the correct crystal pose (\Cref{fig:case-rnp}).
Thus, enhanced TPS reveals how structures, populations, and pathways discovered from a co-folding model must be tested rigorously.

Our framework complements a growing family of methods that sample co-folding models beyond their default predictions.
Two of these methods explore conformational ensembles without training. 
AF3-ReD adds repulsive potentials on the C$\alpha$ RMSD from earlier predictions to the denoising of AlphaFold3, reaching transporter states unresolved at the training cutoff.\citep{ohnuki2025af3red}
Boltz-Jump runs Langevin dynamics on the noised coordinates of Boltz-2 at a single noise level, yielding ensembles five to ten times faster.\citep{daigavane2026boltzjump}
Neither method, unlike enhanced TPS, provides a reweighting scheme to recover the original statistics.
Closer in aim to enhanced TPS are methods that combine a biased diffusion model with sample reweighting.
Feynman--Kac correctors provide the machinery, namely weighted reverse SDEs with sequential Monte Carlo (SMC) resampling.\citep{skreta2025feynman}
Examples include ConforMix, which steers Boltz-1 and BioEmu along the RMSD from their default prediction and doubles the coverage of alternative states in proteins with domain motions,\citep{richman2025conformix}
and enhanced diffusion sampling in BioEmu, which formulates umbrella sampling and metadynamics within these correctors and obtains folding free energies with orders of magnitude fewer samples.\citep{xie2026enhanced}
These methods deliver independent weighted samples, yet their finite-particle approximation precludes exact sampling of the true distribution. 
Enhanced TPS instead builds a Markov chain on the random inputs of the finite-step sampler and targets the output distribution exactly, at the cost of correlated samples.
The approaches are therefore complementary, and enhanced diffusion sampling names the reweighting of path ensembles as a future direction.\citep{xie2026enhanced}

Future efforts will focus on two areas: the evaluation and training of co-folding models.
Although current co-folding benchmarks score single poses, enhanced TPS measures the probability of every state and can thus test whether an ensemble responds to a pocket mutation or a second ligand in line with experiments.
With SuCOS as a collective variable, the same sampling also measures the memorized share of an ensemble (\Cref{sec:experiment}).
Training proximity, equilibrium populations, and structural continuity therefore identify different deficiencies, each of which may be located with enhanced TPS.
We also plan to address post-training, since our case studies indicate that Boltz-2 learns physically plausible conformations with incorrect statistical weights.
By calibrating the model against a classical or machine-learned force field\citep{batatia2026mlff} through on-policy training,\citep{havens2025icml} it may be possible to transform co-folding models into Boltzmann generators\citep{noe2019boltzmann} without an initial large corpus of MD data.
In this part, enhanced TPS would provide such a scheme with the rare states missed by independent generation, together with the weights needed to correct their sampling bias.

Finally, enhanced TPS may help to explain how a diffusion model generates a structure in the first place.
Diffusion models commit to the gross features of a sample within a narrow window of the denoising process, a bifurcation akin to symmetry breaking.\citep{biroli2024regimes}
In molecular simulations, such a commitment is measured by the committor function,\citep{dellago2002tps} which, in the context of co-folding, sets the probability that a partly denoised structure ends in a given state.
BY using enhanced TPS to estimate committor functions, we may test whether transition path theory\citep{e2010transition} can explain how diffusion models work in practice.

\section*{Acknowledgments}
The author acknowledges support by a grant from the Simons Foundation (Grant 839534, MET) and the NYU IT High Performance Computing resources, services, and staff expertise. This work used Jetstream2 at Indiana University and Delta at the National Center for Supercomputing Applications through allocation CHE260231 from the Advanced Cyberinfrastructure Coordination Ecosystem: Services \& Support (ACCESS) program, which is supported by U.S. National Science Foundation grants \#2138259, \#2138286, \#2138307, \#2137603, and \#2138296. Additional computational resources were provided by Mirror Physics through Axon, their agentic platform purpose-built for drug discovery, by Lambda through its Research Grant Program, and by CloudRift.
The author also acknowledges the Google Antigravity Science Program for access to the Gemini 4 Argon coding agent.
The author thanks Antonio Mirarchi for providing the AMARO-v1 checkpoint. 

\section*{Data availability}
The sampler and analysis code is available at \url{https://github.com/muhammadhasyim/tps-cofolding}, and the sampled ensembles are stored on Hugging Face at \url{https://huggingface.co/datasets/muhammadhasyim/tps-cofolding-paper-data}. 

\section*{Competing Interests}
The author declares no competing interests.

\section*{AI use Disclosure}

All code was written using AI coding agents primarily: OpenAI Codex (Codex CLI 0.162.1), Anthropic Claude Code (2.1.322), and Google Antigravity (1.5.10). 
Early code versions were developed in Axon, which combines an AI coding harness with cloud GPU resources for rapid iteration and testing. 
The author specified each algorithm and analysis, reviewed generated code, and validated it against the analytic double well and unit tests (\SICref{si:methods-validation}). 
All figures are drawn from data; no images were generated by AI. 
Agents also assisted with prose, and the author reviewed all text.

\bibliographystyle{unsrtnat}
\bibliography{references}
\end{document}